\documentstyle[11pt,epsf]{article}
\newcommand{\beq}{\begin{equation}}
\newcommand{\eeq}{\end{equation}}
\newcommand{\bea}{\begin{eqnarray}}
\newcommand{\eea}{\end{eqnarray}}
\newcommand{\bi}{\bibitem}
\newcommand{\pslash}[1]{\rlap{/}\kern-0.8pt #1}
\newcommand{\lslash}{\rlap{/}\kern-0.0pt l}
\newcommand{\Dslash}{\rlap{/}\kern-2.0pt D}

\newcommand{\ol}[1]{\overline{#1}}
\newcommand{\loopint}[1]{\int_{-\pi}^\pi {d^4 #1\over(2\pi)^4} }
\newcommand{\momint}[1]{\int_{-\pi/a}^{\pi/a} {d^4 #1\over(2\pi)^4} }

\begin{document}

\title{ 
\vspace{-4.5cm}
\begin{flushright}
{\normalsize BNL--HET--99/2}
\end{flushright}
\vspace{4.0cm}
Calculation of the strange quark mass using domain wall fermions 
}

\author{ {Tom Blum$^{1}$\thanks{Present address: RIKEN BNL Research Center,
        Brookhaven National Lab, Upton, NY 11973, USA},
Amarjit Soni$^1$, and Matthew Wingate$^2$}
\\
$^1${\it Department of Physics, Brookhaven
        National Lab, Upton, NY 11973, USA}\\
$^2${\it RIKEN BNL Research Center,
        Brookhaven National Lab, Upton, NY 11973, USA}
}

\date{\today}
\maketitle

\begin{abstract}

We present a first calculation of the strange
quark mass using domain wall fermions. 
This paper contains an overview of the domain wall discretization
and a pedagogical presentation of the perturbative calculation
necessary for computing the mass renormalization.
We combine the latter with numerical simulations to 
estimate the strange quark mass.  
Our final result in the quenched approximation
is 95(26) MeV in the ${\overline{\rm MS}}$
scheme at a scale of 2 GeV.
We find that domain wall
fermions have a small perturbative mass
renormalization, similar to Wilson quarks, and 
exhibit good scaling behavior.

\end{abstract}

\newpage

\section{Introduction}
\label{sec:intro}

The determination of the quark masses from first
principles is an important task facing particle 
theorists today.  The light quark masses are among the most
poorly determined parameters of the Standard Model.
The cause of the difficulty is the confining
nature of QCD:  quarks exist only in bound states.  Furthermore,
most of the mass of light hadrons is due to the energy of the
color fields surrounding the quarks rather
than the quarks themselves; therefore a nonperturbative
treatment of QCD is required to connect the quark masses
in the QCD Lagrangian with the spectrum of hadronic states
measured experimentally.

Next--to--lowest order
chiral perturbation theory ($\chi$PT) quite precisely predicts the 
ratios of quark masses~\cite{ref:LEUTWYLER} but cannot
set the absolute scale.
The most promising method of computing the light quark masses
(i.e., $m_u$, $m_d$ and $m_s$) is lattice QCD.
It has been suggested that QCD sum rules can be used to 
place fairly strict lower
bounds on $m_s + m_u$ and $m_d + m_u$ by using analyticity
conditions~\cite{ref:LELLOUCH}; however,
calculations of the values of the light quark masses
from sum rules are thought to involve 
many uncertainties~\cite{ref:BGMaltman}.

The feasibility of calculating the quark masses through Monte Carlo
simulation of lattice QCD has been recognized since the 
early days of the field~\cite{ref:FUCINO}.  Most previous work 
utilized two formulations of lattice fermions:
Wilson fermions which explicitly break chiral symmetry at finite
lattice spacing and suffer from large discretization errors; 
and Kogut--Susskind fermions which maintain
a remnant chiral symmetry, but badly break flavor symmetry and seem
to have poorly converged weak--coupling expansions.
Recently, Sheikholeslami--Wohlert (SW) fermions~\cite{ref:SW},
an improvement of Wilson fermions, have also been used to
compute the light quark masses~\cite{ref:ALLTON,ref:FNAL}. 

The usual method of computing the light quark masses on the
lattice is the following.  For fixed gauge coupling and various bare
quark masses, one computes the pseudoscalar
or vector meson mass from the exponential decay in Euclidean time
of an appropriate correlation function.  Using leading order $\chi$PT
one extrapolates in the bare quark mass $m_q$
to where the meson mass takes on its physical value, 
then the corresponding value of the bare quark mass can be 
converted to the renormalized quark mass in any desired scheme.  
This method implicitly uses the vector Ward identity, so the
difference $m_q - m_c$ is renormalized by $Z_m = Z_S^{-1}$, 
where $m_c$ is the bare quark mass corresponding to zero pion mass
and $Z_S$ is
the renormalization constant of the scalar density.
Usually one converts the renormalized lattice quark mass to the continuum
${\ol{\rm MS}}$ regularization scheme by
matching the weak coupling expansions of $Z_S$ 
in both schemes.  Using this procedure
the light quark mass $m_l \equiv (m_u +m_d)/2$ and the 
strange quark mass $m_s$ may be computed independently, for example,
using the pseudoscalar spectrum to fix $m_l$ and the vector spectrum
to fix $m_s$.
A comprehensive analysis of the light quark masses using 
this method appears in Ref.~\cite{ref:LANL}.

Recently several attempts have been made to remove some sources
of uncertainty in the usual method.  A large source of error
in Wilson fermion calculations is the determination of 
the chiral limit; since they explicitly break chiral symmetry,
Wilson quarks become massless at a nonzero critical bare 
quark mass, $m_c \ne 0$.
One can avoid this error by using the axial Ward identity to fix
the bare quark mass~\cite{ref:JLQCD,ref:CPPACS}.  
One computes $\langle 0|\partial_0 A_0|\pi\rangle$ and 
$\langle 0|P|\pi\rangle$, where $A_\mu$ is the local non-singlet
axial vector current and $P$ the non-singlet pseudoscalar density,
and then the quark mass is given by the ratio
$\langle 0|\partial_0 A_0|\pi\rangle/\langle 0|P|\pi\rangle$
and is renormalized by $Z_m = Z_A/Z_P$.  Since the vector meson
and baryon spectra cannot be used with this method, only one
of either $m_l$ or $m_s$ may be fixed independently, the other
is necessarily related by chiral perturbation theory.

Another large uncertainty enters into the matching between
lattice and continuum regularizations.  The typical scale at
which this matching occurs is 2 GeV where the validity
of WCPT is tenuous.  In the case of Kogut--Susskind fermions,
lattice WCPT is untrustworthy: next--to--leading order
corrections can be 50--100\% of the leading order term.  Therefore,
nonperturbative calculation of the renormalization factors
is very desirable.  Two methods are being explored presently
which may remove the need for a perturbative expansion of the lattice 
theory~\cite{ref:PISA},
or push it to a very high energy scale at which the expansion parameter is
much smaller~\cite{ref:ALPHA}.  Finally, the quenched approximation
seems to give quark masses which are roughly 20\% larger than
unquenched quark masses~\cite{ref:LANL}.  Clearly this indicates
that full QCD simulations are necessary for a precise calculation
of light quark masses.

Presently no consensus has been reached regarding the values of the
light quark masses, even within the quenched approximation.  For example,
using Wilson fermions and perturbative matching, the strange quark mass 
is 115(2) MeV when the kaon is used to fix the bare quark mass 
and 143(6) MeV when the $\phi$ meson is used~\cite{ref:CPPACS_LATEST}, 
where the mass is defined in the ${\overline{\rm MS}}$ scheme at 
2 GeV.\footnote{ The widespread belief is that the quenched
approximation yields a $K-\phi$ splitting which is smaller than 
the experimental value.  Using
a regularization independent renormalization scheme, recent
quenched simulations with Kogut--Susskind fermions give
$m_s^{\ol{\rm MS}} = 106(7)$ MeV with the kaon as input versus 
129(12) MeV with the $\phi$ as input~\cite{ref:JLQCDNEWKS}.  
On the other hand, a recent quenched
study using the SW action with a nonperturbatively determined coefficient
claims to reproduce the physical $K-\phi$ splitting if their chiral
fit is quadratic rather than linear~\cite{ref:BBLLMM}.}
Results using the SW action give a lighter strange quark
mass of 95(16) MeV~\cite{ref:FNAL}.  Furthermore, an exploratory
nonperturbative determination of the quark mass renormalization 
agrees with the perturbative renormalization for the usual
quark mass definition, but differs with the perturbative renormalization
for the axial Ward identity definition~\cite{ref:PISA}.
Using the nonperturbative renormalization and the axial Ward identity,
Ref.~\cite{ref:PISA} finds a strange quark mass of 130(18) MeV.
A more comprehensive presentation of the current status appears
in Ref.~\cite{ref:RK}.

In this paper, we employ a new fermion discretization
to compute the light quark masses:  domain wall fermions.
Domain wall fermions utilize a fictitious extra (in this case, 
fifth) dimension
in order to preserve chiral symmetry at nonzero lattice spacing;
the chiral symmetries of the continuum are exactly preserved in the
limit of an infinite fifth dimension~\cite{ref:FURSHAM,ref:KIKUNOGU}. 
The idea originated in the context of chiral gauge theories.  
In Ref.~\cite{ref:KAPLAN}, Kaplan constructed free lattice 
chiral fermions, without doublers, in $2k$ dimensions by considering Dirac
fermions in $2k+1$--dimensions coupled to a mass defect in the
extra dimension, or domain wall. 
For periodic boundary conditions in the extra dimension, an anti--domain wall
also appears which supports $2k$--dimensional chiral fermions of 
the opposite handedness.
Although the suitability of this approach for chiral gauge theories
is still under intensive study, its usefulness
for simulations of
chirally symmetric {\it vector} gauge theories such as QCD
now appears well established
(for a review, see Ref.~\cite{ref:BLUM_LAT98}).

Since the first suggestion that domain wall fermions offer a 
way to study chiral symmetry breaking of 
QCD~\cite{ref:SHAMIR,ref:FURSHAM}, 
considerable work has been done to assess the practicality of the method.
In Ref.~\cite{ref:SHAMIR}, a simplification 
of Kaplan's original proposal
is made for QCD simulations:  half of the extra dimension is discarded,
and the domain 
walls effectively become the boundaries of the extra dimension.
For free field theory
it has been shown that, for a range of the input parameters, 
a light four--dimensional mode of definite chirality is bound to one 
boundary and a similar mode of opposite chirality is bound to the 
other boundary; 
the mixing between the two modes is exponentially 
suppressed with the size of the extra dimension 
\cite{ref:FURSHAM,ref:VRANAS,ref:AOKI,ref:KNY}. 
Significant suppression of the mixing between the modes was also seen
in nonperturbative 
simulations, but whether or not it is purely exponential remains an open 
question~\cite{ref:BLUM_SONI1,ref:BLUM_SONI2,ref:COLUMBIA,ref:BLUM_LAT98}.
Furthermore, the non--singlet axial Ward identity is reproduced, 
and predictions from chiral perturbation theory for the 
dependence of the pseudoscalar meson mass on the quark mass and
for the kaon mixing parameter are
satisfied~\cite{ref:BLUM_SONI1,ref:BLUM_SONI2,ref:BLUM_LAT98}.
Also, the expected behavior of $\langle\bar{q}q\rangle$ 
in the quenched approximation 
due to topological zero modes is 
reproduced~\cite{ref:COLUMBIA,ref:COLUMBIA2}.

The paper is structured as follows:  
Section~\ref{sec:action} introduces the details of the
domain wall fermion action,  Section~\ref{sec:results}
contains the results of the one--loop calculation of
the massive quark self--energy, and Section~\ref{sec:mc} 
gives the details of our Monte Carlo simulations.
In Section~\ref{sec:mq} we combine analytical and numerical
results to give a value for the strange quark mass.
Finally, we present our conclusions in Section~\ref{sec:concl} 
and include some details of our calculation in the Appendix.

\section{Domain wall fermions} 
\label{sec:action}

In this Section we review some properties of domain wall fermions
in QCD, following the original boundary fermion
variant by Shamir~\cite{ref:SHAMIR}.
We take a pedagogical point of view and
introduce notation and methods which will be relevant to
our work later in this paper.  We write down the action and
the propagator, discuss the physics described by the light modes
coupled to the boundaries, and mention previous work supporting
the domain wall formulation of lattice QCD. For further details one should
refer to the literature cited throughout the Section.

\subsection{The action}

On a lattice with spacing $a$, the domain wall fermion action is given by
\beq
- a^4\sum_{x,y}\sum_{s,s'}\bar\psi_s(x/a)D_{s,s'}(x/a,y/a)\psi_{s'}(y/a),
\label{eq:action}
\eeq
where $x,y$ are four--dimensional Euclidean spacetime coordinates and
$s,s'\in[1,N_s]$ are coordinates in the fifth dimension.
The Dirac operator can be separated into a four--dimensional
part, $D^\parallel$, and a one--dimensional part, $D^\perp$:
\beq
aD_{s,s'}(x,y) ~=~ aD^\parallel(x,y) \times \delta_{s,s'} ~+~
 \delta(x-y) \times a D^\perp_{s,s'};
\label{eq:dwfdiracop}
\eeq
$x$ and $y$ have been rescaled to be dimensionless.
The first term is the four--dimensional Wilson--Dirac operator
with a mass term which is negative relative to the usual
$4d$ Wilson fermion action:
\bea
aD^\parallel(x,y) &=& {1\over 2} \sum_\mu \Big[ (1+\gamma_\mu)U_\mu(x)
\delta(x+\hat\mu-y) ~+~ (1-\gamma_\mu)U^\dagger_\mu(y)
\delta(x-\hat\mu-y) \Big] \nonumber \\
& + & (aM - 4)\delta(x-y).
\eea
For $1<s<N_s$, the Dirac operator in the extra dimension is given
by
\beq
aD^\perp_{s,s'} = -\delta_{s,s'} + {1\over2}(1+\gamma_5)\delta_{s+1,s'}
+ {1\over2}(1-\gamma_5)\delta_{s-1,s'}.
\eeq
Note that the five--dimensional fermions are coupled
to {\it four--dimensional} gauge fields which are identical at each
$s$; i.e.\ the link matrices obey
\bea
U_{\mu,s}(x) ~=~ U_\mu(x) &{\rm for}& \mu\in[1,4], \nonumber\\
U_{5,s}(x) ~=~ 1 &{\rm for}& 1 \le s < N_s.
\eea
The boundary conditions in the fifth dimension are anti--periodic
with a weight $am$ which, as we will see, is proportional 
to the $4d$ quark mass.
The Dirac operator for the fifth dimension can
be separated into its chiral components by the projectors,
$P_\pm\equiv(1\pm\gamma_5)/2$, such that
\beq
D^\perp_{s,s'} ~=~ D^{\perp,+}_{s,s'}P_+ ~+~ D^{\perp,-}_{s,s'}P_-.
\eeq
In matrix notation,
\beq
aD^{\perp,+}_{s,s'} ~=~ \left( \begin{array}{ccccc}
	-1 &  1 &  0  & \ldots & 0 \\
	 0 & -1 &  1  & \ldots & 0 \\
	 \vdots  &  \vdots  & \vdots & \vdots & \vdots \\
	 0 &  0 &  0 &  \ldots & 1 \\
	-am & 0 & 0 & \ldots & -1 \end{array} \right) ,
\hspace{1.0cm}
aD^{\perp,-}_{s,s'} ~=~ \left( \begin{array}{ccccc}
	-1 &  0 &  \ldots & 0  & -am \\
	 1 & -1 &  \ldots & 0  & 0 \\
	 0 &  1 &  \ldots & 0  & 0 \\
	 \vdots  &  \vdots  &  \vdots & \vdots & \vdots \\
	 0 & 0 & \ldots & 1  & -1 \end{array} \right).
\label{eq:dperp}
\eeq

Since the perturbative calculation is simpler in (four--)momentum space,
we Fourier transform the ordinary spacetime coordinates.  The
five--dimensional domain wall Dirac operator (\ref{eq:dwfdiracop})
becomes
\bea
aD_{s,s'}(ap) & = & \Big[ \sum_\mu i\gamma_\mu \sin ap_\mu
+ aM - \sum_\mu (1 - \cos ap_\mu)\Big]\delta_{s,s'} + aD^\perp_{s,s'} \\
& = & \sum_\mu i\gamma_\mu a\bar{p}_\mu \delta_{s,s'} + W^+_{s,s'}P_+
 ~+~ W^-_{s,s'}P_-
\label{eq:dwfmomprop}
\eea
where $a\bar{p}_\mu\equiv \sin ap_\mu$, and the $W^\pm$ are
related to the $aD^{\perp,\pm}$ by
\beq
W^\pm_{s,s'} ~=~ aD^{\perp,\pm}_{s,s'} + \Big[aM - \sum_\mu 
(1 - \cos ap_\mu)\Big]\delta_{s,s'}.
\label{eq:wpm}
\eeq

\subsection{The mass matrix}
\label{sec:tree_mass_mat}

The discrete extra dimension
can be interpreted as a flavor space with $N_s - 1$ heavy fermions
and one light flavor~\cite{ref:NN}.  In that framework, the $W^\pm$
in Eqn.~(\ref{eq:dwfmomprop}) are mass matrices which 
govern the flavor mixing.
One can have several physical pictures of how the mass hierarchy
is maintained.
The chiral symmetry is manifest in the domain wall
picture; since the left-- and right--handed modes are bound to opposite
walls in the extra dimension and are separated by a distance $N_s$,
so the chiral components can be rotated independently.  The flavor
space picture also provides insight.  Ref.~\cite{ref:NEU_CHEBY}
relates the mass hierarchy in terms of a generalized see--saw
formula.  In fact, the Froggatt--Nielsen~\cite{ref:FN} mechanism
allows one to establish an approximate conservation law which
protects the light mass from large radiative 
corrections~\cite{ref:NEU_CHEBY}.

Let us examine the eigenvalues and eigenvectors of the
tree--level mass
matrix (in flavor space) for the action described above in
Eqns.~(\ref{eq:action})--(\ref{eq:dperp}).  Details are
presented in Refs.~\cite{ref:SHAMIR,ref:AOKI}, and we repeat
them in Appendix~\ref{sec:diag0} so that they may 
be extended to the one--loop case.
Since the mass matrix is not hermitian, we diagonalize
the mass matrix squared.  Let $\Omega^0$ be the zero momentum
limit of $W^-$~(\ref{eq:wpm}):
\beq
\Omega_0{}_{s,s'} ~=~ \left( \begin{array}{ccccc}
	-b_0 &  0 &  \ldots & 0  & -am \\
	 1 & -b_0 &  \ldots & 0  & 0 \\
	 0 &  1 &  \ldots & 0  & 0 \\
	 \vdots  &  \vdots  &  \vdots & \vdots & \vdots \\
	 0 & 0 & \ldots & 1  & -b_0 \end{array} \right),
\label{eq:omega0}
\eeq
where $b_0 \equiv 1 - M$.  Here and in the rest of the paper,
we rescale $M$ so that it is dimensionless.
As shown in Refs.~\cite{ref:SHAMIR,ref:AOKI,ref:VRANAS}, 
when $|b_0| < 1$ the smallest eigenvalue
of $\Omega_0\Omega_0^\dagger$ (and of $\Omega_0^\dagger\Omega_0$) is
\beq
(\lambda^{(1)})^2 ~=~
(am)^2 M^2 (2-M)^2 + O\Big( (am)^4\Big) + O\Big( (1-M)^{N_s}\Big).
\label{eq:lighteigenval}
\eeq
Therefore, the mass of the light mode, given by $\lambda^{(1)}$,
has an additive renormalization which is suppressed as 
$N_s\rightarrow\infty$ for the range $0 < M < 2$.
(We elaborate on this restriction on $M$ in the next Section and
throughout the remainder of the paper.)
For $|b_0| <1$ the eigenvectors of $\Omega_0\Omega_0^\dagger$ 
are given by
\beq
\phi_s^{(i)} ~=~ \left\{ \begin{array}{ll}
 \sqrt{M(2-M)} ~e^{-\alpha(s-1)}~\Big({\rm sign}~b_0\Big)^{s-1} & i = 1 \\
  \sqrt{ { 2\over N_s}} ~\sin \bigg( {\pi (i-1)\over N_s}
\Big[ N_s + 1 - s\Big]\bigg) & i\ne 1
\end{array}\right.
\label{eq:tree_basis0}
\eeq
where the $i-$th eigenvector corresponds to eigenvalue $(\lambda^{(i)})^2$.
The constant $\alpha_0$ is defined through
\beq
\cosh \alpha_0 ~=~ {1+b_0^2 - (\lambda_0^{(1)})^2\over 2 |b_0|}.
\label{eq:alpha0}
\eeq
Note that the light eigenmode in Eqn.~(\ref{eq:tree_basis0})
is exponentially concentrated at
the $s=1$ boundary while the heavy ($i>1$) modes are not.

Once the eigenvectors of the mass matrix squared have been found,
the Dirac operator can be diagonalized easily.
Following Ref.~\cite{ref:AOKI}
let us define unitary matrices $U^{(0)}$ and $V^{(0)}$ such that
\beq
U^{(0)}_{s,s'} ~\equiv~ \phi_{s'}^{(s)}
\hspace{.5cm}{\rm and}\hspace{.5cm}
V^{(0)}_{s,s'} ~\equiv~ \phi_{N_s+1-s'}^{(s)}.
\label{eq:UnV}
\eeq
Then the basis
\bea
\psi^{\rm diag}_s(p) & \equiv & U^{(0)}_{s,s'}P_+ \psi_{s'}(p) + 
V^{(0)}_{s,s'}P_-\psi_{s'}(p)
\nonumber \\
\bar\psi^{\rm diag}_s(p) & \equiv & \bar\psi_{s'}(p) P_+ 
(V^{(0)}{}^\dagger)_{s',s}
+ \bar\psi_{s'}(p) P_- (U^{(0)}{}^\dagger)_{s',s}
\label{eq:diag0_basis}
\eea
diagonalizes $DD^\dagger$ and $D^\dagger D$.
Furthermore, the Dirac operator itself is diagonal in this basis, 
up to terms which vanish as $N_s \exp(-\alpha_0 N_s)$~\cite{ref:AOKI,ref:KNY}:
\bea
\bar\psi(-p) D(ap) \psi(p) &=& \bar\psi^{\rm diag}(-p)
\Big( V^{(0)} P_+ + U^{(0)} P_- \Big) D \Big( P_+ U^{(0)}{}^\dagger + 
P_- V^{(0)}{}^\dagger\Big)
\psi^{\rm diag}(p) \nonumber \\
&=& \bar\psi^{\rm diag}(-p) \Big( ia\pslash{\bar{p}} ~+~
V^{(0)} W^+ U^{(0)}{}^\dagger P_+ ~+~ U^{(0)} W^- V^{(0)}{}^\dagger P_- \Big) 
\psi^{\rm diag}(p);
\eea
where the $V^{(0)} W^+ U^{(0)}{}^\dagger$ and $U^{(0)} W^-V^{(0)}{}^\dagger$ are 
diagonal in $s,s'$~\cite{ref:AOKI}.
Let us define $\chi\equiv\psi^{\rm diag}_{s=1}$, the eigenstate of
the lightest eigenvalue of the mass matrix (squared).  This 
mode has the effective tree-level action
\beq
{\cal S}^{\rm tree}_{\rm eff} ~=~  a^4\momint{p}~\bar\chi(-p)\Big(i\pslash{p} 
~ + ~m M(2-M)  \Big) \chi(p),
\label{eq:tree_eff_action}
\eeq
since $V^{(0)} W^+ U^{(0)}{}^\dagger\Big|_{s=1,u=1} = 
U^{(0)} W^-V^{(0)}{}^\dagger\Big|_{1,1} =
\lambda_0^{(1)}$~\cite{ref:AOKI}.

\subsection{The propagator}
\label{sec:tree_prop}

The calculation of the tree--level fermion propagator 
$S_F$ proceeds similarly
to the diagonalization of the mass matrix presented above.
The final expression for the propagator
is complicated, so we write it here
schematically; it is written explicitly for the present action
in~\cite{ref:SHAMIR,ref:AOKI} and in Appendix~\ref{sec:rules}.
As in Ref.~\cite{ref:NN} let us write the propagator as
$S_F = D^\dagger /(D D^\dagger)$ and project out its chiral
eigenstates so that
\beq
S^F_{s,s'}(p) ~=~ \Big[(-i\gamma_\mu\bar{p}_\mu\delta_{s,s''} + W^-_{s,s''})
G^R_{s'',s'}P_+
+ (-i\gamma_\mu\bar{p}_\mu\delta_{s,s''} + W^+_{s,s''})G^L_{s'',s'}P_-\Big]
\label{eq:tree_prop2}
\eeq
where $\bar{p}_\mu \equiv \sin p_\mu$, and $G^R$ ($G^L$) is
the inverse of $D^\dagger D$ ($DD^\dagger$):
\beq
G^R_{s'',s'} ~\equiv~ \bigg( {1\over \bar{p}^2 + W^+W^-} \bigg)_{s'',s'}
~~{\rm and }~~
G^L_{s'',s'} ~\equiv~ \bigg( {1\over \bar{p}^2 + W^-W^+} \bigg)_{s'',s'}.
\label{eq:glgr}
\eeq
We give $G^R$ and $G^L$ explicitly in Appendix~\ref{sec:rules},
but let us mention their general behavior.
The homogeneous solutions of 
\beq
\Big(D^\dagger D\Big)_{s,s''} G^R_{s'',s'} = \delta_{s,s'}
~~{\rm and }~~
\Big(D D^\dagger\Big)_{s,s''} G^L_{s'',s'} = \delta_{s,s'}
\label{eq:diffeq4gs}
\eeq
are exponentials: $\exp(\pm\alpha(p) s)$.  If $0<b(p)<1$ then
the solutions are decaying exponentials;
$\alpha(p)$ is real and defined through
\beq
\cosh \alpha(p) ~=~ {\bar{p}^2 + 1 + b^2(p)\over 2|b(p)|},
\label{eq:coshalpha}
\eeq
where
\beq
b(p) ~\equiv~ 1 - M + \sum_\mu (1 - \cos ap_\mu).
\label{eq:b}
\eeq
If $-1<b(p)<0$ then the solutions oscillate:
$(-1)^s\exp(\pm\alpha(p) s)$, with $\alpha$
as in Eqn.~(\ref{eq:coshalpha}).
For $|b(p)|>1$ there is no longer a
mode bound to the domain wall: the solutions of Eqn.~(\ref{eq:diffeq4gs})
go like $\exp(\pm i\alpha(p) s)$ where $\alpha$ is now defined
through
\beq
\cos \alpha(p) ~=~ {\bar{p}^2 + 1 + b^2(p)\over 2b(p)},
\label{eq:cosalpha}
\eeq
Thus, $M$ must be in the range
\beq
0 < M < 2
\label{eq:dwheight}
\eeq
in order for there to be a single massless fermion as $N_s\rightarrow\infty$.
At greater values of $M$, ``doubler'' states in the 
other corners of the Brillouin zone become nearly massless.
For example, the four states with one
component of momentum near $\pi/a$ contribute, and so on in increments 
of 2 up to $8<M<10$ where again only a single state with all four 
momenta near $\pi/a$ exists. In fact the action is symmetric under the
changes $M\to 10-M$ and $\psi_s(x)\to(-1)^{\sum_\mu x_\mu+s}\psi_s(x)$
so the physics of any
region and the one reflected about $M=5$ are identical~\cite{ref:YIGAL}.
Henceforth, we concentrate our discussion on the single flavor theory
near the origin of the Brillouin zone.
Also note that the range  of $M$ in Eqn.~(\ref{eq:dwheight})
is additively renormalized in the interacting theory, and
we discuss this renormalization in detail in Sec.~\ref{sec:Mcrit}.

\subsection{Summary of tree level properties}
\label{sec:tree_sum}

Let us emphasize the following points.
For strictly infinite $N_s$ (and $m=0$) there
is a massless right--handed fermion bound to the $s=1$ 
wall and a left--handed fermion bound to the anti--domain wall at
$s=\infty$~\cite{ref:KAPLAN}.  For large, but finite, $N_s$ 
the mixing between the two chiralities is exponentially 
suppressed at tree level~\cite{ref:SHAMIR}.
One describes a light 4--component Dirac fermion 
by coupling the two light modes with an explicit chiral symmetry
breaking term $m$~\cite{ref:FURSHAM}.
Then the light Dirac fermion has a mass in the free 
theory equal to~\cite{ref:SHAMIR,ref:VRANAS}
\beq
am_q^{(0)} ~=~ M(2-M)\Big[ am ~+~ (1 - M)^{N_s}\Big].
\label{eq:mq0}
\eeq
Therefore, neglecting exponentially small terms,
domain wall fermions describe a light mode whose mass is 
{\it multiplicatively} renormalized. 
It also turns out that the light mode satisfies 
continuum--like axial Ward identities~\cite{ref:FURSHAM}.
These features make domain wall fermions very attractive
for simulating light quark physics, where chiral symmetry is
crucial.


Another virtue of the domain wall formulation is that 
in the limit $N_s\to\infty$ the
leading  discretization errors for the $4d$ effective action are $O(a^2)$.
In the massless theory,
any gauge invariant dimension--five operator with the required lattice
symmetries can be written as a linear combination of the following two 
operators~\cite{ref:SW}:
\bea
O_1 &=& \bar{q} ~D^2~ q \\
O_2 &=& {i\over2}\bar{q} ~\sigma_{\mu\nu} F_{\mu\nu}~ q 
\eea
where $D^2$ is the second--order covariant derivative
and
$F_{\mu\nu}=[D_\mu,D_\nu]$ is the field strength tensor.
However, neither of these terms is invariant under
a chiral transformation
\bea
q &\rightarrow& e^{i\epsilon\gamma_5} q \nonumber \\
\bar{q} &\rightarrow& \bar{q} e^{i\epsilon\gamma_5}. 
\eea
Since chirality violating effects have been shown to vanish
 as $N_s\to \infty$, the contributions of
$O_1$ and $O_2$ to the effective action must be suppressed. 
In this sense, the domain wall fermion action is an
$O(a)$--improved action~\cite{ref:KNN,ref:BLUM_SONI2}.
Even with an $O(a)$-improved action, $O(a)$ errors can enter 
into observables as the operators may require
$O(a)$ improvement as well.  However, such improvements would
also violate chiral symmetry and, by the same argument as above,
are suppressed as $N_s\to\infty$.
Of course precise scaling tests
are necessary to evaluate the extent of the improvement;
although simulations to date~\cite{ref:BLUM_SONI2},
including the ones in this work, are consistent with these expectations.

\section{Perturbative mass renormalization}
\label{sec:results}

In this Section we present our one--loop calculation of the quark
mass renormalization.  A renormalization factor is needed in order
to match a lattice definition of quark mass to a continuum definition.
We are also pursuing methods of computing renormalization factors 
nonperturbatively; however, that work is beyond the scope of this paper.  

We compute the matching factor $Z_m(\mu,a)$ between a quark mass 
defined on a lattice with
spacing $a$ to a continuum quark mass renormalized at momentum scale $\mu$:
\beq
m^{\overline{\rm MS}}(\mu) ~=~ Z_m(\mu,a)~m^{\rm LAT}(a).
\label{eq:mcontmlat}
\eeq
$Z_m$ is computed by equating the one--loop continuum fermion propagator 
to the one--loop lattice fermion propagator.
In this Section we present the calculation of the full five--dimensional 
self--energy, and then we discuss its effect on $M$ and $m$.
The wavefunction renormalization has been computed already in
Ref.~\cite{ref:AOKI}; we have extended that work to the massive case
and present the full one--loop calculation here for clarity.\footnote{
	While this manuscript was in preparation, an independent
	one--loop calculation of the quark mass renormalization 
	appeared in Ref.~\cite{ref:AIKT} which helped us in tracking
	down an error in our preliminary work~\cite{ref:WINGATE_LAT98}.
	}
Only the main points are made in the body of the Section, while
more details are given in Appendix~\ref{sec:details}.

\subsection{Five--dimensional fermion self--energy}
\label{sec:5dimse}

\begin{figure}
\vspace{1.0in}
\includegraphics{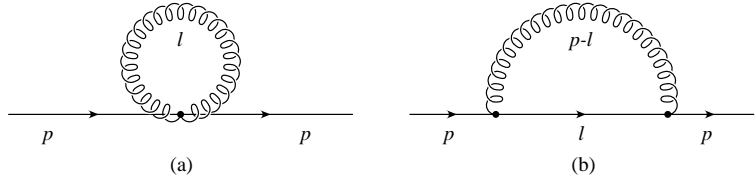}
\caption{ One--loop fermion self--energy diagrams: the (a) tadpole and
(b) half--circle graphs.}
\label{fig:loops}
\end{figure}

The fermion self--energy, $\Sigma(p,m)$, is given to one--loop order
by the Feynman diagrams shown in Figure~\ref{fig:loops}.   
We use $p$ to denote the external momentum and $l$ to denote the
momentum in the loop integral.
The tadpole graph has no fermion propagator in the loop, so it
has trivial dependence on the fifth dimension, i.e.\ it is diagonal.
On the other hand, the fermion 
in the loop of the half--circle graph may propagate in the fifth
dimension (change flavor) while the gluon is unaffected.  
Therefore the half--circle graph has off--diagonal contributions in
$s,s'$ space.

Even with the extra dimension, the steps of evaluating the
half--circle graph are much like those for the calculation using
Wilson fermions \cite{ref:GONZ,ref:BSD}.  First an integral which
has the same infrared ($p\rightarrow 0$) limit is subtracted from
$\Sigma(p,m)$ to cancel logarithmic divergences.  The difference
may then be Taylor expanded about zero lattice spacing.  
In the continuum limit, we neglect
terms in the expansion which vanish as $a\rightarrow 0$; since
the coefficients of $O(a)$ terms are exponentially suppressed with
increasing $N_s$ the leading discretization errors are, in
effect, $O(a^2)$.  The resulting expression can be arranged as follows:
\beq
\Sigma ~=~ {g^2 C_F\over 16\pi^2} \Big[ {1\over a}\Sigma_0 + 
i\pslash{p}\Sigma_1 + m\Sigma_2\Big].
\eeq
After a lengthy calculation which is presented in Appendix~\ref{sec:details},
these terms can be further subdivided into
\bea
\Sigma^{(0)}_{s,s'} & = & I^{(0)}_{s,s'} ~-~ 2T\delta_{s,s'} 
\label{eq:sigma0} \\
\Sigma^{(1)}_{s,s'} & = & L^{(1)}_{s,s'} ~+~ I^{(1)}_{s,s'} 
~-~ {T\over 2} \delta_{s,s'}
\label{eq:sigma1} \\
\Sigma^{(2)}_{s,s'} & = & L^{(2)}_{s,s'} ~+~ I^{(2)}_{s,s'}
\label{eq:sigma2} 
\eea
where the $L$ terms are proportional to $\ln a$, and the $I$ and
$T$ terms are finite integrals to be computed numerically.  
We only need the renormalization of the lightest mode, so we
delay further evaluation until we rotate the terms to the basis
which diagonalizes the one--loop mass matrix.
We should note, however, that the $I$ and $L$ terms are functions of $M$.
Since $M$ becomes additively renormalized, this dependence is
the source of a systematic uncertainty: what numerical value
of $M$ should one use to compute $I^{(1)}$ and $I^{(2)}$?
We will address this issue in Sec.~\ref{sec:Mcrit}.
To summarize, the five-dimensional one--loop effective action is
given by
\bea
\bar\psi_s(-p)\bigg\{ & a^{-1} &\bigg[W^+P_+ ~+~ W^-P_-
~+~ {g^2 C_F\over 16\pi^2} \Big(I^{(0)} - 2T\Big) \bigg] \nonumber \\
&+&i\pslash{p}\bigg[ 1 + {g^2 C_F\over 16\pi^2}
 \bigg(L^{(1)} + I^{(1)} - {T\over 2}\bigg) \bigg]
 + m {g^2 C_F\over 16 \pi^2} \Big( L^{(2)} + I^{(2)} \Big) \bigg\}_{s,s'}
\psi_{s'}(p) .
\label{eq:5dim_eff}
\eea

\subsection{Renormalization of $M$}
\label{sec:Mcrit}

We noted above (\ref{eq:dwheight}) that in the free theory
$M=0$ is the critical point where light modes begin to appear
bound to the domain walls.
The $1/a$ contribution from the self--energy graphs,
$\Sigma_0$ given in Eqn.~(\ref{eq:sigma0}), shifts
this value in the same way 
that the massless limit of Wilson--like fermions is shifted:
\beq
M = 0 ~\to~ M = M_c,
\eeq
where $M_c$ is the point at which the domain wall action first 
supports light chiral modes on the boundaries.
If the tadpole contribution is dominant the shifts are identical since
that graph is the same for domain wall and Wilson fermions.

Of course, the whole range of $M$ corresponding to light modes is 
renormalized when
the coupling is non--zero, except that $M=5$ is a fixed point
since the action is still symmetric under $M\to 10-M$ and 
$\psi_s(x)\to(-1)^{\sum_\mu x_\mu+s}\psi_s(x)$. 
For example, for numerical simulations it is helpful to know 
the optimal value of $M$ which minimizes the extent of the light mode in the
fifth dimension. At tree--level the wavefunction
corresponding to the zero mode bound to the $s=1$ domain wall
is a $\delta$--function in the $s$--direction when $M=1$; i.e., $M=1$
is the optimal value for free domain wall fermions since the intrinsic
quark mass arising from mixing
of modes on opposite walls is minimized.
However, simulations at $\beta\approx 6.0$ have shown that the 
optimal $M$ at which axial symmetries are preserved is 
somewhere around $1.7$ and that 
axial symmetries are poorly respected at $M=1.0$
\cite{ref:BLUM_SONI1,ref:BLUM_SONI2,ref:BLUM_LAT98}.

The question now becomes, how is the whole range of $M$ renormalized? 
It cannot be a simple uniform shift since $M=5$ is a fixed point. 
If we consider only the tadpole contribution to the
self--energy, the shift is approximately uniform in each region 
($0 < M < 2$, $2< M< 4$, $\ldots$), weighted by a factor 
\bea
\sum_{\mu}(1-\cos{(p_\mu)})\to 4,2,0,-2,-4
\eea
coming from the Wilson term. Surprisingly, this simple picture also
describes the nonperturbative data well, as will be shown later in
this paper.
This tadpole--improved
estimate of $M$ was originally proposed in Ref.~\cite{ref:AOKI}.

Perturbatively, the shift of $M_c$ is given through
the $a^{-1}$ terms in Eqn.~(\ref{eq:5dim_eff}):
\beq
M_c ~=~ -{g^2 C_F\over 16\pi^2}\Sigma^{(0)}
 ~=~ -{g^2 C_F\over 16\pi^2} \Big(I^{(0)} - 2T\Big).
\label{eq:mc1loop}
\eeq
$I^{(0)} = I^{(0)}_{s,s'}$ is not diagonal in the extra
dimension, the one--loop calculation of $M_c$ for the light mode 
involves rotating to the basis which diagonalizes the one--loop
mass matrix.\footnote{The calculation has
been carried through in Ref.~\cite{ref:AIKT}.}  However, a reasonable
first estimate for $M_c$ assumes the tadpole graph is numerically
much larger than the contribution from $I^{(0)}$:
\beq
M_c^{\rm tad} ~\equiv~ {2 g^2 C_F T\over 16\pi^2},
\label{eq:mctad}
\eeq
where $T$ is given in Eqn.~(\ref{eq:T}) and is numerically
equal to 24.4.  In Table~\ref{tab:Mcrit} we give two
values of $M_c^{\rm tad}$, each computed with different
definitions of the strong coupling constant, $g^2_V(3.41/a)$
and $g^2_V(1/a)$.  There is an obvious problem in deciding the
relevant scale of this effect.
In Section~\ref{sec:mq} we discuss the choice of coupling constant
and scale in detail, but it is clear that a perturbative 
estimate of $M_c$ is not precise enough for our purposes. We note, however,
that for a reasonable estimate of the renormalized coupling, the shift
due to the tadpole agrees well with the above nonperturbative estimate.

\begin{table}
\caption{\label{tab:Mcrit} Values of $M_{\rm crit}$
estimated from the tadpole graph and computed with the
4--d Wilson fermion action.}
\begin{center}
\begin{tabular}{l|ccc}
& $\beta = 6.3$ & $\beta=6.0$ & $\beta=5.85$ \\ \hline \hline 
$M_c^{\rm tad}(3.41/a)$ & 0.676 & 0.754 & 0.812  \\ 
$M_c^{\rm tad}(1/a)$ & 0.985 & 1.17 & 1.41\\ \hline
$\kappa_c^W$\cite{ref:BBS} & 0.1519 & 0.1572 & 0.1617\\
$M_c^W$ & 0.708 & 0.819 & 0.908
\end{tabular}
\end{center}
\end{table}

Due to the above considerations, it is desirable to have a 
nonperturbative determination
of $M_c$ since it appears in the definition of the lattice quark mass. 
A direct search for $M_c$ would be numerically prohibitive, so 
it is fortunate that there is a simple nonperturbative estimate available
which originates from 
the overlap description of domain wall fermions~\cite{ref:NN_OL}
(thus it is exact only for $N_s\to\infty$).
In this case a transfer matrix can be defined which describes propagation in
the fifth dimension ${\cal T}=\exp(\gamma_5 H^W(-M))$ where
\beq
H^W(-M) ~\equiv~ \gamma_5 D^\parallel(-M)
\eeq
is the ordinary $4d$ Hermitian Wilson--Dirac Hamiltonian with a mass
term that is negative of the conventional one.
$D^\parallel$ first supports exact zero modes 
as $M$ approaches a critical value $M_c^W$ defined by a vanishing pion mass.
In this case ${\cal T}$ has a unit eigenvalue and propagation 
in the fifth dimension
is unsuppressed. Thus $M_c$ for domain wall fermions corresponds
to $M_c^W$ for Wilson fermions, usually given in terms of the
hopping parameter, $\kappa_c^W$:
\beq
M_c^W ~=~ -\Bigg({1\over 2\kappa_c^W} - 4\Bigg).
\label{eq:kappac2m}
\eeq
We can simply take $M_c=M_c^W$ from existing numerical simulations.
In Table~\ref{tab:Mcrit} we give the values of $\kappa_c^W$
computed in Ref.~\cite{ref:BBS} and the corresponding 
value of $M_c^W$ which we use for the rest of this work.  

\begin{figure}
\vspace{3.5in}
\includegraphics{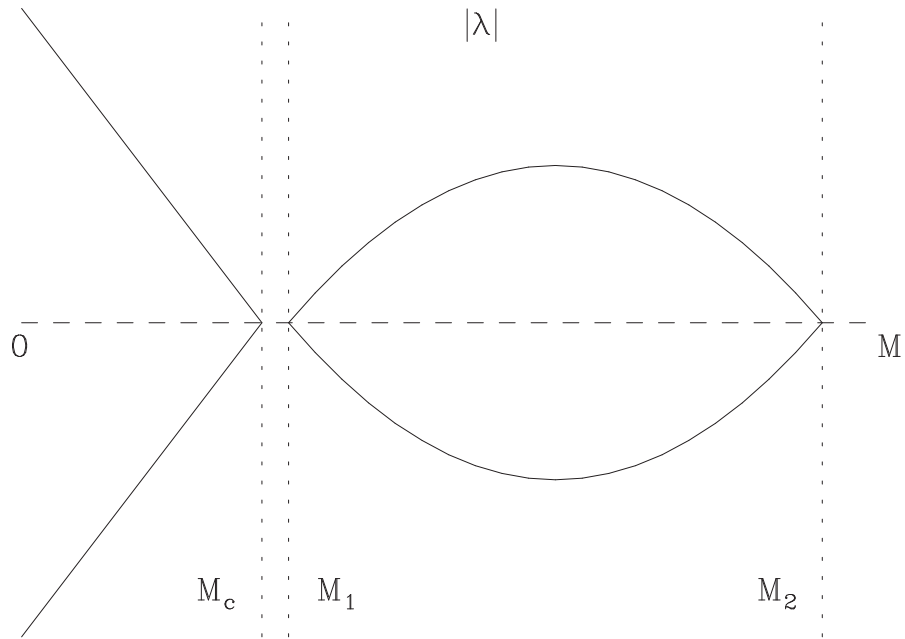}
\caption{ Expected spectrum of the Wilson--Dirac operator $D^\parallel$
as a function of $M$ (which is negative compared to the usual
convention for Wilson fermions).  There is a mass gap for $M<M_c$
and $M_1 < M < M_2$, and no gap in the Aoki region $M_c<M<M_1$.
}
\label{fig:gap}
\end{figure}

In Figure~\ref{fig:gap} we show schematically what the spectrum
of the four--dimensional Wilson--Dirac operator should look like
(with the domain wall convention for the sign of $M$) for the single
flavor case.
Ordinary Wilson fermion simulations are performed in the region $M < M_c$.
while the domain wall fermion simulations have $M_1< M < M_2$.
It has been conjectured by Aoki~\cite{ref:AOKI_PHASE} that there is a range
$M_c < M < M_1$ where there is no gap due to the spontaneous breaking of
flavor and parity; evidence for the existence of this Aoki phase has been 
found in lattice QCD simulations~\cite{ref:NUMERICAL_AOKI} and
analytically using an effective chiral Lagrangian~\cite{ref:SS}
where the width of this phase, $M_1-M_c$,
was found to be $O(a^3)$. In the conventional picture (see Ref.~\cite{ref:SS}
for a recent summary) the gap reopens after $M_1$ and closes again at
$M_2$ when four of the doubler modes become massless.

The spectrum of the Hermitian Wilson--Dirac Hamiltonian has been studied
in some detail in Ref.~\cite{ref:EHN}. Many
level crossings in the Hamiltonian were found uniformly in
the region above $M_1$. These zeroes were related to small instanton-like
configurations and are presumably related to the same 
lattice artifacts that give
rise to so--called exceptional configurations~\cite{ref:BDEHT}.
If the density of these zero modes is non--zero in the large volume limit, then
the gap is closed and domain wall fermions cannot exist:
the entire allowed region of $M$ should then be in the Aoki phase. Since
numerical 
evidence~\cite{ref:BLUM_SONI1,ref:BLUM_SONI2,ref:COLUMBIA,ref:COLUMBIA2}
 for the existence of domain wall fermions is quite
strong, we infer that the density of these zeroes vanishes in
the large volume limit (see Ref.~\cite{ref:BLUM_LAT98} for more details),
at least for the couplings relevant to present simulations. This is also 
consistent with the above studies of the Aoki phase. However, we also note that
at some strong coupling the conventional picture is for the whole region to
be in the Aoki phase, and at this coupling domain wall fermions cease to exist.
The existence of the Aoki phase also reveals why the size of the extra dimension
must increase as the coupling becomes stronger: as the gap gets smaller 
the size of the extra dimension must increase to maintain
the same amount of suppression.
We refer the reader to Ref.~\cite{ref:NEU_CHEBY} for similar plausibility
arguments on the behavior of domain fermions at strong and weak coupling.

In closing this section we emphasize that the simple replacement of
\beq
M ~\rightarrow~ \tilde{M} ~\equiv~ M - M_c^W
\label{eq:mtilde}
\eeq
in the quark mass (\ref{eq:mq0}),
though nonperturbative, is an ansatz which may or may not introduce 
$O(a)$ errors.  
Furthermore, $M$ cannot really be
uniformly shifted, even piecewise, over the whole region. Nonperturbative
effects such as instanton--like artifacts may be important 
(though these do seem
to be more or less uniformly distributed above $M_c$).
However, as previously mentioned,
Eqn.~(\ref{eq:mtilde}) is a very good fit to our numerical data which
we present in Sec.~\ref{sec:mc}.
Also the identification in Eqn.~(\ref{eq:kappac2m}) is only exact in the 
limit $N_s\to\infty$.
Again, simulations indicate this is a good 
approximation (see Ref.~\cite{ref:BLUM_LAT98}).

\subsection{Quark mass renormalization}
\label{sec:z_m}

In this Section we concentrate on the renormalization of
the quark mass.  We follow the method outlined in the case
of the wavefunction renormalization~\cite{ref:AOKI}.
The tree level quark mass was given
in Sec.~\ref{sec:tree_mass_mat} by finding the 
smallest
eigenvalue of the mass matrix squared, $\Omega_0\Omega_0^\dagger$.
At one--loop level, the mass matrix is renormalized:
\beq
\Omega ~\equiv~ \Omega_0 + g^2 \Omega_1.
\eeq
$\Omega_0$ is the tree--level mass matrix (\ref{eq:omega0}) and
$\Omega_1$ is the one--loop correction given by the terms
in (\ref{eq:5dim_eff})
\beq
\Omega_1 ~\equiv~ {C_F\over16\pi^2}\Big(\Sigma^{(0),-} + 
am\Sigma^{(2),-}\Big).
\eeq
where $\Sigma^{(j),\pm}\equiv\Sigma^{(j)} P_\pm$.

\begin{figure}
\vspace{3.5in}
\includegraphics{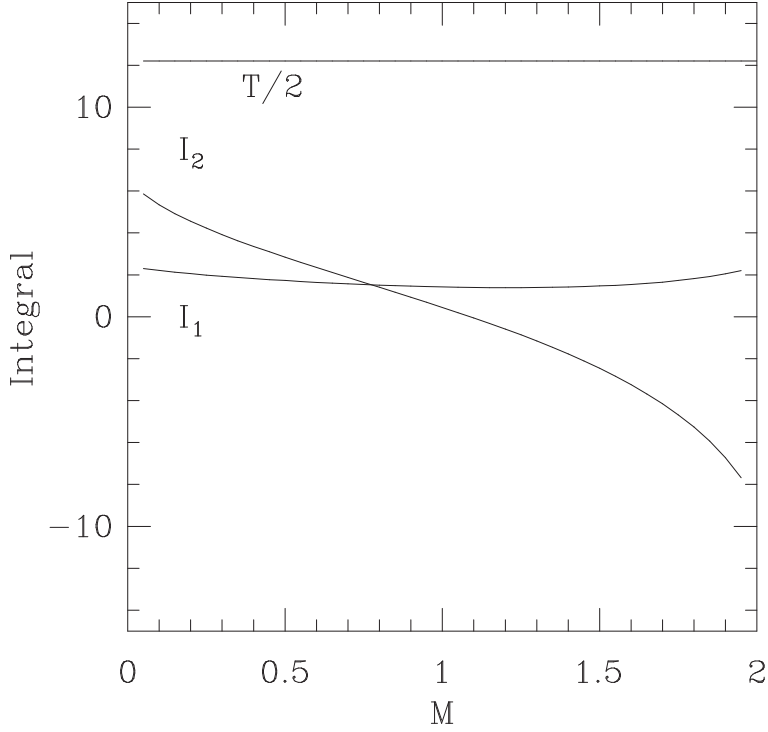}
\caption{ The three finite integrals evaluated at values of $M$,
(see Eqns.~(\ref{eq:I1})--(\ref{eq:T})).
}
\label{fig:s_diag}
\end{figure}

We rotate the fermion fields to the basis which diagonalizes the
one--loop matrix $\Omega$
\bea
\psi^{\rm diag}_s(p) & \equiv & U_{s,s'}P_+ \psi_{s'}(p) + 
V_{s,s'}P_-\psi_{s'}(p)
\nonumber \\
\bar\psi^{\rm diag}_s(p) & \equiv & \bar\psi_{s'}(p) P_+ (V^\dagger)_{s',s}
+ \bar\psi_{s'}(p) P_- (U^\dagger)_{s',s}.
\label{eq:diag_basis}
\eea
Then the terms which control the renormalization of the light fermion
mode are as follows:
\bea
V W^+ U^\dagger \Big|_{s=1,u=1} &= U W^- V^\dagger \Big|_{1,1} ~=~ &
mM(2-M)
\\
V I^{(0),+} U^\dagger \Big|_{1,1} &= U I^{(0),-} V^\dagger \Big|_{1,1} ~=~ &
O(N_s b_0^{N_s}) ~\rightarrow~ {\rm negligible}\\
U L^{(1),+} U^\dagger \Big|_{1,1} &= V L^{(1),-} V^\dagger \Big|_{1,1} 
~\equiv~ &\tilde{L}_1
\\
U I^{(1),+} U^\dagger \Big|_{1,1} &= V I^{(1),-} V^\dagger \Big|_{1,1} ~\equiv~&
\tilde{I}_1 \\
V L^{(2),+} U^\dagger \Big|_{1,1} &= U L^{(2),-} V^\dagger \Big|_{1,1} 
~\equiv~& M(2-M)~\tilde{L}_2 
\\
V I^{(2),+} U^\dagger \Big|_{1,1} &= U I^{(2),-} V^\dagger \Big|_{1,1} ~\equiv~&
M(2-M) ~\tilde{I}_2 .
\eea  
The results of Ref.~\cite{ref:AOKI} for $\tilde{L}_1$ and $\tilde{I}_1$
combined with our results for $\tilde{L}_2$ and $\tilde{I}_2$ are that
\bea
\tilde{L}_1 & =& -2\int_0^1 dx ~x\ln \Big({\pi\over az}\Big)^2 ~+~ {3\over 2}
\label{eq:L1}
\\
\tilde{L}_2 & =& -4\int_0^1 dx ~\ln \Big({\pi\over az}\Big)^2 ~+~ 4
\label{eq:L2}
\\
\tilde{I}_1 & = & -16\pi^2\loopint{l}~ \Bigg\{ {1\over 8 \hat{l}^2} 
\sum_\mu \bigg[
\sin^2l_\mu (\tilde{G}_R + \tilde{G}_L) + 2 \cos l_\mu (b_0 - b(l))
\tilde{G}_R\Big] \nonumber \\
&+& \sum_\mu {\sin^2 l_\mu\over 2 \hat{l}^4}\Big[(b_0-b(l))\tilde{G}_R
- \Big( 2\cos^2{l_\mu\over2} - \sum_\nu \cos^2{l_\nu\over2}\Big)\tilde{G}_L
+ \hat{l}^2 \tilde{G}_R \Big] \Bigg\}
\nonumber \\
& + & 16\pi^2\loopint{l} ~\Theta(\pi^2 - l^2) ~{1\over l^4}
\label{eq:I1}
\\
\tilde{I}_2 & = & -16\pi^2 \loopint{l}~ {1\over \hat{l}^2} 
{1\over(1 - e^{-\alpha(l)} b_0)^2}
~ \Bigg[ \bar{l}^2 \sum_\mu \cos^2{l_\mu\over2}
{1\over(1 - e^{\alpha(l)} b(l))^2} - \bar{l}^2{e^{-\alpha(l)}
\over(1-e^{\alpha(l)}b(l))}
 \nonumber \\
& + &  \sum_\mu\sin{l_\mu\over2} e^{-2\alpha(l)} \Bigg]
~+~ 4(16\pi^2)\loopint{l} ~\Theta(\pi^2 - l^2) ~{1\over l^4}
\label{eq:I2} \\
T &=& 16\pi^2 \loopint{l}~ {1\over \hat{l}^2}.
\label{eq:T}
\eea
where $z^2 = (1-x)(p^2 x + (m_q^{(0)})^2)$,
$\bar{l}_\mu \equiv \sin l_\mu$ and
$\hat{l}_\mu \equiv 2\sin(l_\mu/2)$.  We have also used the definitions
\beq
\tilde{G}_R(l)  ~=~ {1\over 2b(l)\sinh\alpha(l)}\bigg[ {(b_0^{-1}-e^{-\alpha(l)}) 
- (b_0-e^{\alpha(l)})
\over (b_0^{-1}-e^{-\alpha(l)}) + (b_0-e^{\alpha(l)})} ~-~ {1-b_0^2\over
(e^{\alpha(l)} - b_0)^2} \bigg],
\eeq
and
\beq
\tilde{G}_L(l) ~=~ {1\over 2b(l)\sinh\alpha(l)}\bigg[ {(b_0^{-1}-e^{-\alpha(l)}) 
- (b_0-e^{\alpha(l)})
\over (b_0^{-1}-e^{-\alpha(l)}) + (b_0-e^{\alpha(l)})} ~-~ {1-b_0^2\over
(e^{\alpha(l)} - b_0)^2}{e^{\alpha(l)} - b(l) \over e^{-\alpha(l)}-b(l)} \bigg].
\eeq
The quantities $b(l)$, $b_0=b(0)$, and $\alpha(l)$ are
defined in Sec.~\ref{sec:tree_prop}
The finite integrals $\tilde{I}_1, \tilde{I}_2$, and $T$ are
plotted as functions the five--dimensional
mass $M$ in Figure~\ref{fig:s_diag}.

\begin{figure}
\vspace{3.5in}
\includegraphics{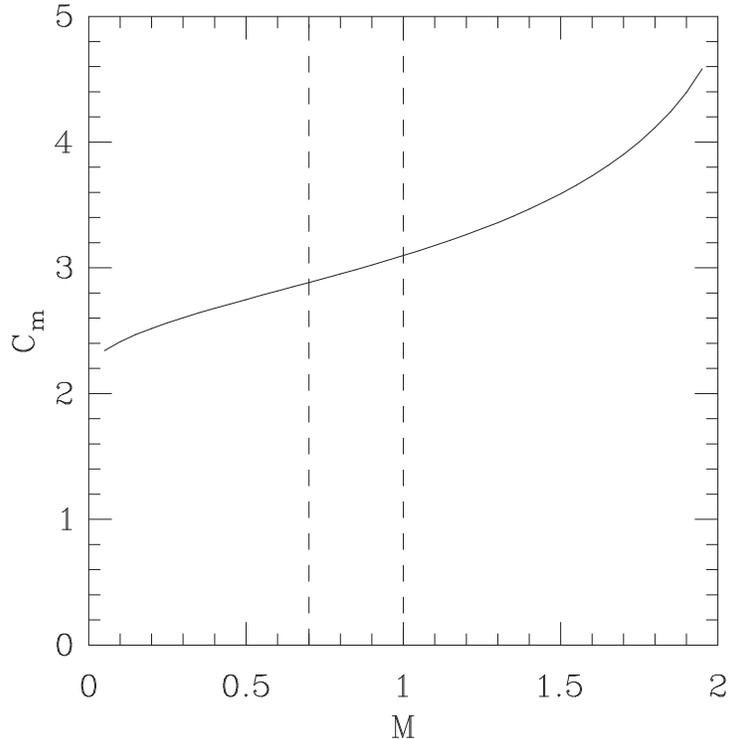}
\caption{ The matching coefficient $C_m$ as a function $M$.
The dashed lines indicate the range of $M$ where numerical
simulations have been performed. }
\label{fig:c_m}
\end{figure}

The expression for the fermion self--energy in the continuum, using
dimensional regularization in the $\overline{\rm MS}$ scheme, can
be written as
\beq
\Sigma(p,m_f) ~=~ i\pslash{p}\Sigma^{\overline{\rm MS}}_1
~+~ m_f \Sigma^{\overline{\rm MS}}_2,
\eeq
where $m_f$ is the continuum mass and
\bea
\Sigma_1^{\ol{\rm MS}} &=& {2}\int_0^1 dx ~x 
~\ln\Big({Q^2\over z_c^2}\Big) - 1, \\
\Sigma_2^{\ol{\rm MS}} &=& {4}\int_0^1 dx
~\ln\Big({Q^2\over z_c^2}\Big) - 2,
\eea
and $z_c^2 \equiv (1-x)(p^2 x + m_f^2)$.  
The lattice mass and
continuum mass are matched onto each other by combining these
calculations so that, in Eqn.~(\ref{eq:mcontmlat}),
\beq
Z_m ~=~ {Z^{\rm LAT}_2 \over Z^{\rm LAT}_1} \times
{Z^{\overline{\rm MS}}_1 \over Z^{\overline{\rm MS}}_2},
\eeq
with the notation that $Z_j \equiv 1 - g^2 C_F \Sigma_j /16\pi^2$.
The final mass renormalization factor between domain wall fermions
and continuum fermions in the $\overline{\rm MS}$ scheme is given by
\beq
Z_m ~=~ 1 ~-~ {6 g^2 C_F\over 16\pi^2}\bigg( \ln(\mu a) - C_m \bigg),
\label{eq:zm}
\eeq
where
\beq
C_m ~=~ \ln\pi - {1\over 4} + {1\over6}\bigg({T\over2} + \tilde{I}_1 - 
\tilde{I}_2\bigg).
\eeq
As discussed in detail in Sec.~\ref{sec:Mcrit}, 
the parameter $M$ is renormalized.    
%
Since the perturbative estimate of the renormalization~(\ref{eq:mc1loop})
is large and untrustworthy, we use the ansatz~(\ref{eq:mtilde}).
The integrals (\ref{eq:I1}) and (\ref{eq:I2}) are unchanged,
except the trivial replacement $M\to\tilde{M}$ in Figs.~\ref{fig:s_diag}
and \ref{fig:c_m}.
We mark the region of $\tilde{M}$ where our simulations have been performed
with vertical dashed lines.
Between those, $0.7 < \tilde{M} < 1.0$, $C_m$ varies from 2.88 to 3.10.  
These values of $C_m$ should be compared to $C_m=2.16$ for Wilson fermions,
3.22 for Sheikholeslami--Wohlert fermions, and 6.54 for Kogut--Susskind
fermions.

\section{Monte Carlo results}
\label{sec:mc}

In practice,
Monte Carlo simulation of quenched QCD using domain wall fermions
is very similar
to standard calculations using Wilson fermions.  We use the
conjugate gradient algorithm to invert the five--dimensional
fermion matrix.  Next, $4d$ quark fields are constructed from the
$5d$ fields at the two boundaries~\cite{ref:FURSHAM}:
\bea
q(x) &=& P_+ \psi_1(x) ~+~ P_- \psi_{N_s}(x) \nonumber \\
\bar{q}(x) &=& \bar\psi_{N_s}(x) P_+ ~+~ \bar\psi_1(x) P_-
\eea
These are the simplest interpolating fields for the lightest mode, and 
composite
operators constructed from them satisfy exact continuum--like 
Ward identities in the limit $N_s\to\infty$~\cite{ref:FURSHAM}.

For this exploratory calculation of the strange quark mass, we compute
the pseudoscalar meson mass and decay constant on a few dozen
configurations at three lattice spacings.  Specifically we perform
Monte Carlo simulations at 
three values of the gauge coupling: $\beta \equiv 6/g^2 = $
5.85, 6.0, and 6.3.  The size of the extra dimension for the main
part of this work for the three couplings was $N_s = $ 14, 10, and
10, respectively, and the $5d$ mass parameter was $M = $ 1.7, 1.7, and
1.5, respectively. 
The value of $M \approx 1.7$ at 6.0 is the optimal value which suppresses
propagation of the light mode in the extra dimension, as found in a previous
study~\cite{ref:BLUM_SONI2}. The $M$ at the other two $\beta$'s
were {\it ad hoc} choices
based on the $\beta=6.0$ value and the fact that the optimal $M$ should 
decrease to 1 in the weak coupling limit.
At all three gauge couplings the mesonic two--point functions were computed
with $m = $ 0.075 and 0.050, and at $\beta = 6.0$ and 6.3
another mass, $m = 0.025$, was also included.  The lattice volumes
at $\beta = 6.0$ and 6.3 are roughly $\approx$(1.6 fm)${}^3$
while the $\beta = 5.85$ volume is $\approx$(2.0 fm)${}^3$.
Our raw lattice simulation results are given in Table~\ref{tab:data1}.

\begin{table}
\caption{\label{tab:data1} Pseudoscalar mass and decay constant for
the main numerical data set.}
\begin{center}
\begin{tabular}{c|cc|cc|cc}
 & \multicolumn{2}{c}{$\beta=5.85, 16^3\times 32$} 
 & \multicolumn{2}{c}{$\beta=6.0, 16^3\times 32$} 
 & \multicolumn{2}{c}{$\beta=6.3, 24^3\times 60$} \\
$m$ & $aM_\pi$ & $af_\pi$  & $aM_\pi$ & $af_\pi$ 
& $aM_\pi$ & $af_\pi$\\ \hline \hline
0.025 & -- & -- & 0.309(6) & 0.076(4) &  0.245(5) &  0.056(5) \\
0.050 & 0.488(5) & 0.104(4) & 0.423(5) & 0.088(4) &  0.340(4) &  0.066(3) \\
0.075 & 0.588(4) & 0.114(4) & 0.517(5) & 0.094(4) &  0.425(4) &  0.072(3) 
\end{tabular}
\end{center}
\end{table}

In addition to the gauge coupling and bare quark mass, 
domain wall fermion simulations depend on the five--dimensional
mass $M$ and the size of the extra dimension $N_s$.
The study of
how data are affected by changing these parameters is important.
Given the results of Ref.~\cite{ref:BLUM_SONI2}, we believe
that we have taken $N_s$ large enough so that corrections due to
the mixing of the two light modes in the center of the extra dimension
are smaller than the rest of our uncertainties which we will show later
to be $\approx 10\%-20\%$.  
The effect of a finite $N_s$ is to give the pion an additional
intrinsic mass $m_I$:
\beq
(aM_\pi)^2 ~=~ A~\tilde{M}~(2-\tilde{M})~(m ~+~ m_I).
\eeq
In the free theory $m_I = |1-M|^{N_s}$~\cite{ref:VRANAS}, 
and in the interacting theory $m_I$ is expected to decay exponentially
$m_I \sim \exp(-\alpha N_s)$.

\begin{figure}
\vspace{3.5in}
\includegraphics{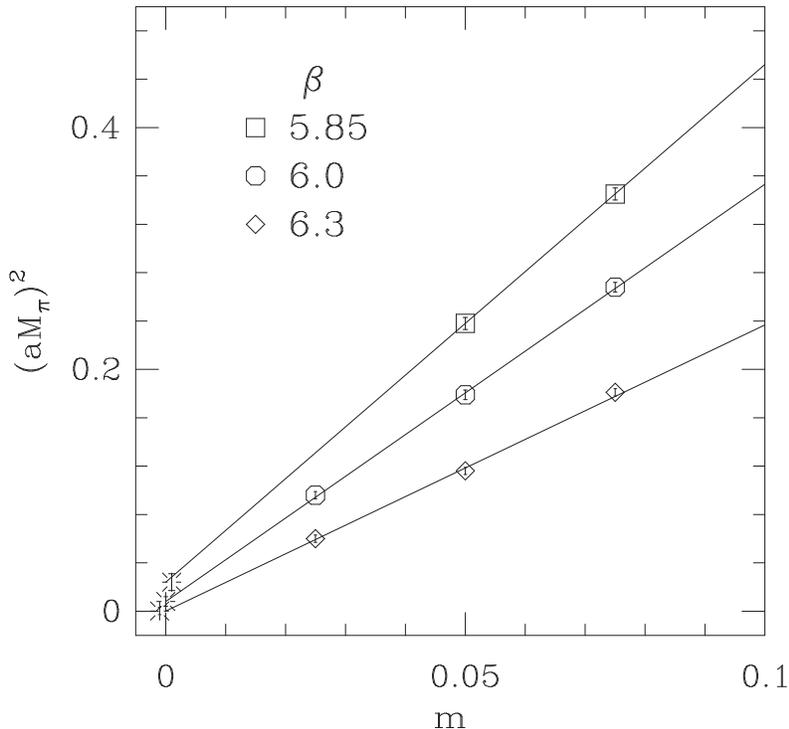}
\caption{ Lattice pseudoscalar mass vs.\ $m$ for the three
$\beta$ values.  Asterisks indicate linear extrapolations to $m=0$
with jackknifed errors and are slightly displaced horizontally for clarity.}
\label{fig:mpi2_m_beta}
\end{figure}

Indeed we find a nonzero intercept
when we extrapolate the pion mass squared to $m=0$ for $\beta = 5.85$
and 6.0, (see Fig.~\ref{fig:mpi2_m_beta}), but find that the $\beta=6.3$
mass squared does extrapolate to zero. 
We have repeated simulations at $\beta = 6.0$ using $N_s = 14$
and found the pion mass decreases by a few percent (compare
Tables~\ref{tab:data1} and \ref{tab:data2}) signaling a 
statistically significant $m_I$.  However, errors in determining
the lattice spacing from $f_\pi$ are much larger than the difference
in $aM_\pi$ and so the $N_s = 10$ results are sufficient for this
work.  

A thorough study of the $N_s$ dependence at a stronger coupling,
$\beta=5.7$, was presented in Ref.~\cite{ref:COLUMBIA,ref:COLUMBIA2}.
They find two important results. First, physical results were unchanged
between $N_s=32$ and 48. This is consistent with arguments above concerning
the behavior of domain wall fermions at stronger coupling. They used $M=1.65$
in their study. It is possible that a larger value would decrease the value
of $N_s$ required to reach the asymptotic region. Second, even in
this limit, the pion mass does not vanish as $m\to 0$, which is then presumably
a ($4d$) finite volume effect.

We were able to test for finite volume errors at $\beta = 6.0$
by computing the pion mass with $N_s=10$ on a 
$24^3\times 40$ lattice.  The results, $aM_\pi = 0.318(5)$ and
0.427(4), for $m = 0.025$ and 0.050 respectively, are not statistically
distinguishable from the results on the $16^3\times 32$ lattice 
(see Tab.~\ref{tab:data1}).

\begin{table}
\caption{\label{tab:data2} Pseudoscalar mass $(aM_\pi)$ for the $\beta = 6.0$,
$N_s = 14$ numerical data set.}
\begin{center}
\begin{tabular}{c|cccc}
$m$ & $M=1.5$ & $M=1.7$ & $M=1.9$ & $M=2.1$ \\ \hline \hline
0.025 & 0.290(9) & 0.293(7) & 0.290(7) &  0.281(11)\\
0.050 & 0.394(7) & 0.411(5) & 0.412(7) & 0.396(8) \\
0.075 & 0.487(7) & -- & 0.510(7) & 0.490(7)
\end{tabular}
\end{center}
\end{table}

\begin{figure}
\vspace{3.5in}
\includegraphics{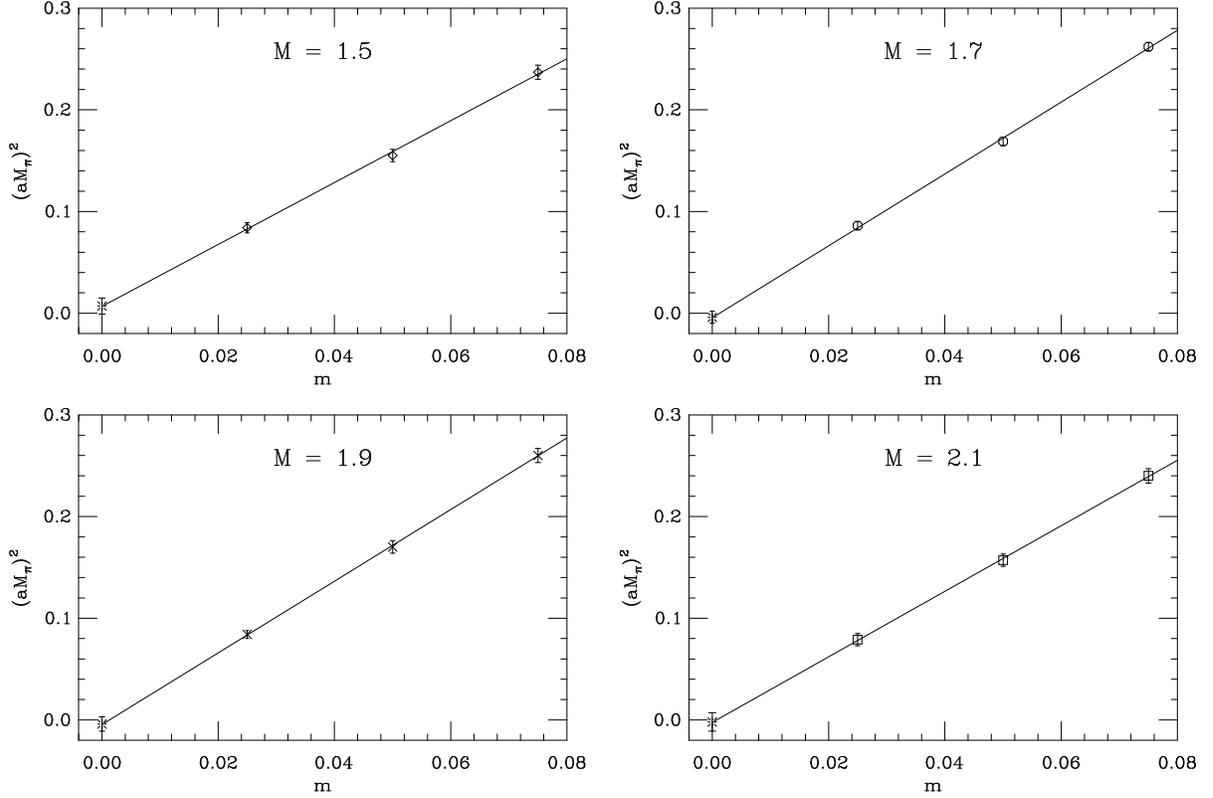}
\caption{ Pion meson mass squared as a function of $m$ for $\beta=6.0$.  
Lines are least squares fits which extrapolate to $(aM_\pi)^2=0$ at $m=0$
for all four values of $M$ within the errors shown by asterisks.}
\label{fig:mpi2_m_4}
\end{figure}

\begin{figure}
\vspace{3.5in}
\includegraphics{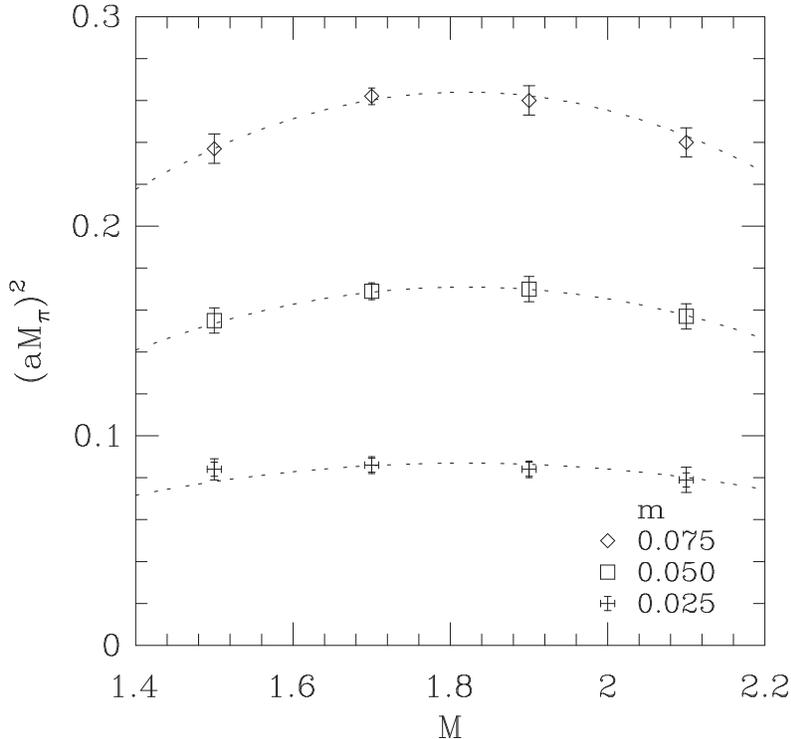}
\caption{ Pion meson mass squared as a function of $M$ at
$\beta = 6.0$.  $N_s=14$ except for the $M=1.7$, $m=0.075$ point
which has $N_s=10$.  The dotted lines are fits to
$(aM_\pi)^2 = Am\tilde{M}(2 - \tilde{M})$. }
\label{fig:mpi2_m0}
\end{figure}

Since the quark mass, even at tree level, explicitly depends
on $M$, we performed several more runs at $\beta=6.0$ with $m=0.025, 0.050$
and 0.075, varying $M$ between
1.5 and 2.1 on roughly 20 configurations with $N_s=14$.\footnote{
In the interest of frugality, we did not extend the $m=0.075, M=1.7$ data
from $N_s=10$ to $N_s=14$.}  Table~\ref{tab:data2} displays the
values of $aM_\pi$ obtained from these simulations, and
Fig.~\ref{fig:mpi2_m_4} shows $(aM_\pi)^2$ as a function of $m$.  
The linear least squares
fits extrapolate to $(aM_\pi)^2=0$ at $m=0$ within the (uncorrelated)
errors for each $M$ separately.  
Furthermore, in Fig.~\ref{fig:mpi2_m0}
we plot $(aM_\pi)^2$ as a function of $M$ for
the three values of $m$.  
We can test the ansatz~(\ref{eq:mtilde}) simply by fitting
the data to
\beq
(aM_\pi)^2 = A~m~\tilde{M}~(2-\tilde{M}),
\label{eq:mpivsmtilde}
\eeq
for each $m$.
The dashed lines in Fig.~\ref{fig:mpi2_m0} are fits to 
Eqn.~(\ref{eq:mpivsmtilde}) 
and have good $\chi^2$'s; each $\chi^2$ per degree of freedom is
less than one, and $A$ is the same within (uncorrelated) errors for each $m$:
$A=$ 3.48(8), 3.42(6), and 3.52(4) for $m=$ 0.025, 0.05, and 0.075,
respectively.
Therefore, the ansatz (\ref{eq:mtilde}) is well justified in
this work.
At present the data do not permit a more general fit.
Given these observations
we find the definition of the lattice quark mass
\beq
m^{\rm LAT} ~=~ m~\tilde{M}~(2-\tilde{M})
\label{eq:latmass}
\eeq
to be very reasonable and suitable for this work.

\begin{figure}
\vspace{3.5in}
\includegraphics{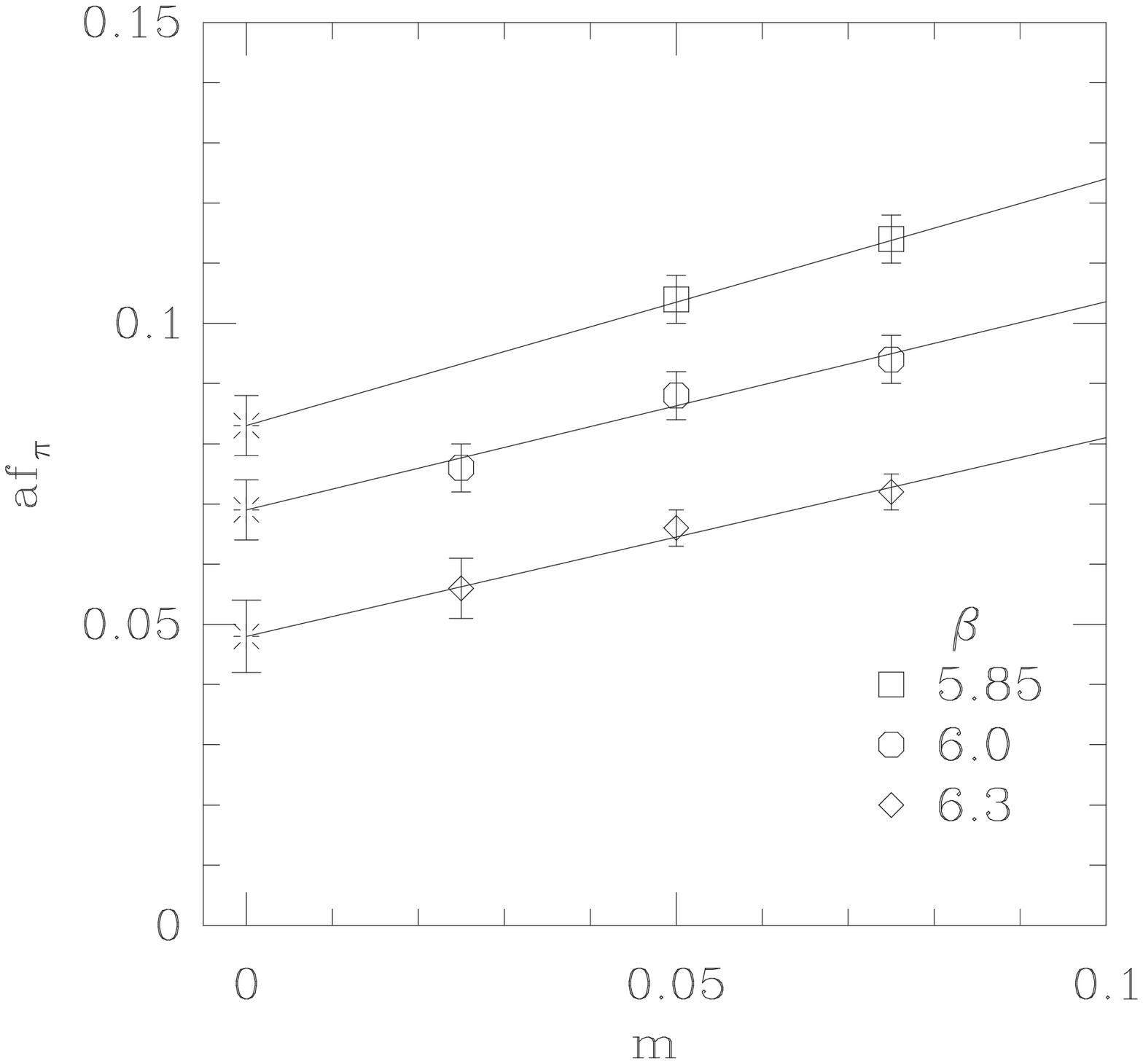}
\caption{ Lattice pseudoscalar decay constant vs.\ $m$ for the three
$\beta$ values.  Asterisks indicate linear extrapolations to $m=0$
with jackknifed errors. }
\label{fig:fpi_m}
\end{figure}

\section{Quark mass and coupling constant}
\label{sec:mq}

We believe we have a good understanding of the dependence of hadron
spectrum on $M$, so now we concentrate on the data at one
value of $M$ per lattice spacing (as listed in Table~\ref{tab:mcresults}).
We compute the pion decay constant $a f_\pi$ for three masses 
$m$ for each $\beta$ and extrapolate linearly to the chiral limit, $m=0$,
(see Fig.~\ref{fig:fpi_m}).
We determine the lattice spacing by setting this extrapolated
value to the physical pion decay constant, $f_\pi = 130.7$ MeV.
Here we must emphasize that in this exploratory calculation, the
determination of the inverse lattice spacing has a large
systematic uncertainty due to having a only a few data points
to extrapolate $af_\pi$ to $m=0$.  
In the continuum, $\chi$PT gives a one--loop correction to
the decay constant which goes as~\cite{ref:LANG_PAGELS}
\beq
f_\pi ~=~ f \bigg[ 1 + {M_\pi^2\over (4\pi f)^2} \ln \bigg( 
{M_\pi^2 \over \Lambda^2} \bigg) \bigg],
\eeq
where $f$ is the tree--level decay constant and $\Lambda$ is the
cutoff.  With only three quark masses we cannot resolve the
logarithmic behavior of $f_\pi$, so we extrapolate linearly to $m=0$,
(see Fig.~\ref{fig:fpi_m}).
We should remark that the signal for the decay constant is rather noisy:
$af_\pi$ varies by as much as 10\% depending on the range of Euclidean
time over which one computes correlation functions.
Since the determination of the lattice spacing for this work is through
the extrapolation of $af_\pi$ to $m=0$, it comes with a large uncertainty.


\begin{table}
\caption{\label{tab:mcresults} Summary of simulation parameters and
results.}
\begin{center}
\begin{tabular}{l|ccc}
& $\beta = 5.85$ & $\beta=6.0$ & $\beta=6.3$ \\ \hline \hline
\# configs. & 18 & 30 & 11 \\
volume & $16^3 \times 32$ & $16^3 \times 32$ & $24^3 \times 60$  \\
$N_s$  & 14 & 10 & 10 \\
$M$ & 1.7 & 1.7 & 1.5 \\
$a^{-1}(f_\pi)$ (GeV) & 1.57(15) & 1.89(14) & 2.72(34) \\
$m_s$(MeV) & 73(7) & 75(6) & 76(10) \\
$M_c^W$ & 0.908 & 0.819 & 0.708 \\
$C_m$ & 2.94 & 3.01 & 2.94 \\
$m_s^{\rm LAT}$(MeV) & 70(7) & 74(6) & 73(10) \\
$\langle{\rm Tr}~U_{\rm plaq}/3\rangle$ & 0.5751 & 0.5937 & 0.6224 \\
$\alpha_V(3.41/a)$ & 0.157 & 0.146 & 0.131 \\
$Z_m(\mu = 2{\rm GeV})$ & 1.30 & 1.34 & 1.34 
\end{tabular}
\end{center}
\end{table}

Next, we use chiral perturbation theory as a guide to interpolate
the pion mass squared linearly in the quark mass: $M_\pi^2 \sim m$
to the value of $m$ which gives the physical kaon mass,
495 MeV.  
That value must then be doubled to take into account the
fact that we use degenerate quarks in our simulations, while the
physical kaon has one strange quark and one lighter non--strange
quark.\footnote{To be accurate one should use $m = (m_s+m_l)/2$, 
setting $m_l$ with the physical $M_\pi$.  However, in this exploratory
work, we have neglected this $\approx 4\%$ effect.}
We denote the parameter $m$ corresponding to the physical 
kaon by $m_s$ and display our results in Table~\ref{tab:mcresults}.
Using Eqn.~(\ref{eq:latmass}) we obtain $m_s^{\rm LAT}$, 
the domain wall strange quark mass.


Combining Eqns.~(\ref{eq:mcontmlat}) and (\ref{eq:zm})
the following expression relates a quark mass
computed on a lattice with spacing $a$
to the quark mass defined in the modified minimal subtraction
scheme of dimensional regularization at momentum scale $\mu$:
\beq
m_s^{\overline{\rm MS}}(\mu) ~=~
m_s^{\rm LAT}\Bigg[ 1 - {2\alpha_s\over\pi}\bigg( \ln(\mu a) 
- C_m \bigg)\Bigg],
\label{eq:matchmq}
\eeq
where we have substituted $g^2/4\pi = \alpha_s$ and $C_F=4/3$.
The last quantity we need is the coupling constant $\alpha_s$.
In a one--loop calculation such as this, there is 
ambiguity in the definition of the coupling constant.
The matching equation (\ref{eq:matchmq}) is derived by
equating the poles of the one--loop quark propagators
computed in the continuum and on the lattice.  Each procedure
uses a differently defined coupling constant; however,
the difference between the two coupling constants is a higher order
correction in perturbation theory:
\beq
\alpha_s^{\rm LAT} - \alpha_s^{\rm CONT} ~=~ O(\alpha_s^2).
\eeq

It has been known for some time that the bare coupling
constant is not a good expansion parameter for lattice 
perturbation theory~\cite{ref:LM}.
Therefore we use a physical definition related to the heavy
quark potential.  Specifically, we define the coupling constant
using the two--loop perturbative expression for the plaquette 
(the $1\times 1$ Wilson loop):
\beq
-\ln\bigg\langle{1\over3}{\rm Tr}~U_{\rm plaq}\bigg\rangle ~=~
{4\pi\over 3}\alpha_V(3.41/a)\Big[1 ~-~ 1.19\alpha_V
~+~ O(\alpha_V^2)\Big].
\label{eq:alphavdef}
\eeq
The scale $3.41/a$ has been computed by estimating and minimizing
the effect of higher order terms~\cite{ref:LM}.  One can run
$\alpha_V$ to any other scale using the universal two--loop
beta function.
Alternatively, one can compute the continuum $\overline{\rm MS}$
coupling constant $\alpha_{\overline{\rm MS}}$ from $\alpha_V$
(at a scale $q$) perturbatively~\cite{ref:BLM}:
\beq
\alpha_{\overline{\rm MS}}(q e^{-5/6}) ~=~
\alpha_V(q)\Big[1 + {2\alpha_V\over \pi} + O(\alpha_V^2)\Big].
\eeq
Although the three--loop beta function has been computed for
$\alpha_{\overline{\rm MS}}$~\cite{ref:RODRIGO}, we choose to
use the two--loop beta function for consistency; the difference
comes in below the 1\% level.

The remaining problem is to decide on the scales and corresponding
couplings to insert in the matching equation (\ref{eq:matchmq}).
Refs.~\cite{ref:JI,ref:GBS} advocate the reorganizing of
lattice perturbation theory as we described above, so that
the resulting expression may be ``horizontally matched''
to the continuum perturbative expansion.
In converting our lattice quark mass into a continuum quark mass
we follow the procedure described in Ref.~\cite{ref:LANL}.
Then the continuum matching scale $\mu$ should be set to
the ``best'' lattice scale which minimizes the higher order
corrections to the fermion self--energy.  Unfortunately, it is
harder to estimate $\mu$ for logarithmically divergent graphs
than it was in the case of the plaquette.  Therefore we resort
to trying a spread of $\mu$ values.  Evidence from previous work
indicates that in the range $\mu = 0.5/a$ to $\pi/a$ the 
higher order, ultraviolet--dominated, effects are 
minimized~\cite{ref:LM,ref:LANL}.
Finally, the quark mass is run to 2 GeV using the two--loop running 
equation~\cite{ref:RUNNINGMASS}.  We also test the systematic
error by repeating the procedure for different values of $\mu$.
In Table~\ref{tab:masses} we give the strange quark masses 
at the matching scales, $m(\mu)$, and the mass run to $\mu=2$ GeV;
we also give our values of $\alpha_{\overline{\rm MS}}$ 
at different scales.   Our statistical errors are 3--4 times larger
than the variation in $m$(2 GeV) computed with different values of $\mu$,
so we cannot argue that one particular scale minimizes higher
order effects in Eqn.~(\ref{eq:matchmq}).  

\begin{table}
\caption{\label{tab:masses} Strange quark masses (in MeV) 
in the $\overline{\rm MS}$ scheme at increasing values of the 
matching scale $\mu$.  We give the mass at both the matching scale
$\mu$ and at 2 GeV.  The first pair of masses are without
tadpole improvement and the second pair include tadpole
improvement. }
\begin{center}
\begin{tabular}{cc|cc|cc}
$\mu$ & $\alpha_{\overline{\rm MS}}(\mu)$ & $m(\mu)$ & $m$(2 GeV) &
$m^{\rm TI}(\mu)$ & $m^{\rm TI}$(2 GeV) \\ \hline
\multicolumn{6}{c}{ $\beta = 5.85$ } \\ \hline
$0.5/a$ & 0.275 & 114(11) & 96(9) & 108(11) & 90(9) \\ 
$1/a$ & 0.199 & 96(9) & 92(9) & 93(9) & 90(9) \\ 
$2.0$ GeV & 0.181 & 91(9) & 91(9) & 90(9) & 90(9) \\ 
$2/a$ & 0.156 & 85(8) & 90(9) & 85(8) & 90(9) \\ 
$\pi/a$ & 0.137 & 81(8) & 90(9) & 81(8) & 90(9) \\ 
\hline 
\multicolumn{6}{c}{ $\beta = 6.0$ } \\ \hline
$0.5/a$ & 0.243 & 116(9) & 102(8) & 111(8) & 97(7) \\ 
$1/a$ & 0.182 & 99(8) & 99(8) & 97(7) & 96(7) \\ 
$2.0$ GeV& 0.178 & 98(8) & 98(8) & 96(7) & 96(7) \\ 
$2/a$ & 0.146 & 89(7) & 97(7) & 89(7) & 97(7) \\ 
$\pi/a$ & 0.129 & 85(6) & 96(7) & 85(7) & 97(7) \\ 
\hline
\multicolumn{6}{c}{ $\beta = 6.3$ } \\ \hline
$0.5/a$ & 0.202 & 107(14) & 101(13) & 103(14) & 97(13) \\ 
$2.0$ GeV& 0.175 & 99(13) & 99(13) & 97(13) & 97(13) \\ 
$1/a$ & 0.159 & 94(12) & 98(13) & 93(12) & 96(13) \\ 
$2/a$ & 0.131 & 86(11) & 97(13) & 86(11) & 97(13) \\ 
$\pi/a$ & 0.118 & 82(11) & 97(13) & 83(11) & 97(13) 
\end{tabular}
\end{center}
\end{table}

One might try to improve the perturbation expansion 
by nonperturbatively estimating ultraviolet effects which spoil
the convergence.  For example, the tadpole--improvement prescription
advocates perturbatively expanding the quantity $m^{\rm LAT}/u_0$,
where $u_0$ is designed to be sensitive to short--distance 
fluctuations~\cite{ref:LM}.
Therefore, the tadpole--improved quark mass, in the $\overline{\rm MS}$
scheme is related to $m^{\rm LAT}/u_0$ perturbatively by
\beq
m^{\rm TI}_{\overline{\rm MS}}(\mu) ~=~ \Big(m^{\rm LAT}\Big/u_0\Big)
\bigg[ 1 - {2\alpha_s\over\pi}\bigg( \ln \mu a ~-~
\Big[C_m - C_T\Big]\bigg)\bigg],
\eeq
where $u_0$ is defined to be the fourth root of the plaquette
\beq
u_0 ~\equiv~ \bigg\langle {1\over3} {\rm Tr}~U_{\rm plaq}
\bigg\rangle^{1/4},
\eeq
and it is computed from Monte Carlo simulation.
The matching coefficient is modified at the one
loop level by the perturbative expansion of $u_0$
\beq
u_0 ~=~ 1 - {1\over12}g^2  ~\equiv~ 1 - {2\over\pi}C_T\alpha_s,
\eeq
which defines $C_T$.
In Table~\ref{tab:masses} we give the tadpole--improved strange
quark mass for the various matching scales, $m^{\rm TI}(\mu)$,
as well as the mass run to 2 GeV, $m^{\rm TI}$(2 GeV).
At the lower scales $\mu \le 2$ GeV, tadpole--improvement lowers the mass
by 4--6 MeV indicating that the unimproved perturbative result has
significant higher order corrections.  On the other hand, there is
no significant difference between the standard and improved expansions
when the matching is done at $\mu > 2$ GeV.
Therefore, we will choose for our final result
the improved mass with the matching
done at $\mu=2/a$, $m_{\overline{\rm MS}}^{\rm TI}(2/a)$ in
Table~\ref{tab:masses}, and assign a 2 MeV systematic error
due to the arbitrary choice of scale.  We perturbatively 
run our final result to 2 GeV for comparison with other results.

\begin{figure}
\vspace{3.5in}
\includegraphics{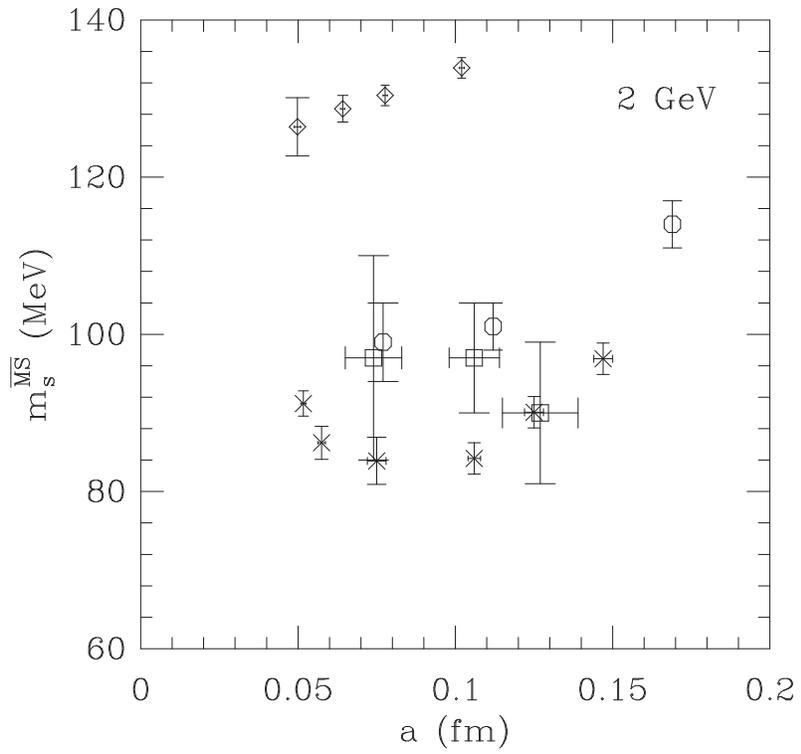}
\caption{ Strange quark mass in $\overline{\rm MS}$ scheme at 2 GeV. 
Our results are displayed
as squares, Wilson fermions as diamonds~\cite{ref:CPPACS_LATEST},
Sheikholeslami--Wohlert fermions as circles~\cite{ref:FNAL},
and Kogut--Susskind results as crosses~\cite{ref:JLQCD_STAG,ref:KIM_OHTA}.}
\label{fig:mstrange_msbar}
\end{figure}

In Figure~\ref{fig:mstrange_msbar} we plot our results for
the strange quark mass (with statistical error bars only)
in the $\overline{\rm MS}$ scheme at the scale 2 GeV 
 along with those obtained using other fermion discretizations.
In choosing the data to which we compare the domain wall results,
we used those which were obtained with the same method.
The masses were computed by fixing the kaon to be its physical
mass and the matchings to the continuum were computed perturbatively.
Unfortunately, we can only set the lattice spacing using $f_\pi$
while the other data in Fig.~\ref{fig:mstrange_msbar} use the
$\rho$ meson mass.  In typical Wilson fermion 
simulations~\cite{ref:GF11}, as well
as preliminary domain wall fermion simulations~\cite{ref:BS_JERUSALEM},
there is a $\approx 15\%$ systematic uncertainty due to choosing 
$f_\pi$ vs.\ $M_\rho$ to set the scale.

In this first work, the statistical errors are rather large, about 15\%, 
yet it is encouraging that we see no significant signs of scaling
violations.  As argued in Sec.~\ref{sec:tree_prop}, discretization errors 
should be $O(a^2)$ rather than $O(a)$ since wrong--chirality operators 
are suppressed; however, a more precise study is needed to draw a
firm conclusion.  Since no scale dependence can be detected within
the statistical errors in this exploratory study, we take a 
weighted average of the strange quark mass determined at the three
lattice spacings to give 95 MeV with a purely statistical uncertainty
of 5 MeV.  A linear extrapolation to the continuum limit raises
the central value by 15 MeV, but of course has a large uncertainty.
Given that we do not expect to have $O(a)$ scaling violations,
we take 95 MeV as our strange quark mass but add the 15 MeV
in quadrature with the rest of our systematic errors.

The largest systematic error arises from the ambiguity in which
physical quantity is used to fix the lattice spacing.  As mentioned
above, preliminary results and experience with Wilson fermions 
lead us to believe there could be at least a 15\% uncertainty using
domain wall fermions.  Considering that
our calculation of $f_\pi$ was done with only a few quark
masses, we conservatively estimate a 20\% systematic
uncertainty in $a^{-1}$;
the other systematic errors are small in comparison.  
For example, the uncertainty in matching scale induces an error
of 2 MeV.  We have not yet computed the strange quark mass by
fixing the $\phi$ meson to its physical mass, but in Wilson fermion
simulations there is an $O(10\%)$ difference from the strange quark
mass using the kaon.  For now we include this error as a systematic
uncertainty in our calculation, but it must be explicitly checked
with domain wall fermions.  
Of course quenching also induces a systematic error; 
however, we are unable to address its effects without
a full dynamical--fermion simulation.

\begin{table}[t]
\caption{\label{tab:syserr} Sources of systematic uncertainty in computing
the strange quark mass.}
\begin{center}
\begin{tabular}{l|c}
Source & Estimated size \\ \hline
Using $f_\pi$ vs.\ $M_\rho$ to set $a$ & 20\% \\
Matching scale, $\mu$ & 2\% \\
Using $\phi$ vs.\ $K$ to set $m_s$ & 10\% \\
Continuum extrapolation & 16\%
\end{tabular}
\end{center}
\end{table}

Adding the statistical and systematic errors in quadrature,
our final result for the strange quark mass within the quenched
approximation is
\beq
m_{\overline{\rm MS}}(2 ~{\rm GeV}) ~=~ 95(26) ~{\rm MeV}.
\label{eq:ta_da}
\eeq

\section{Conclusions}
\label{sec:concl}

This work has focused on the steps needed to compute the
light quark masses with the domain wall fermion discretization.
We have extended the calculation of the domain wall fermion
self--energy to the massive case.  We find that the 
perturbative mass renormalization
factor which matches the domain wall lattice regularization
to the $\overline{\rm MS}$ regularization scheme is as well--behaved
as that for Wilson fermions. 

In conjunction with the perturbative calculation, we have
performed numerical simulations of quenched lattice QCD using
domain wall fermions.  At $\beta = 6.0$ the pion mass squared
vanishes linearly in $m$ as $m \rightarrow 0$, and the ansatz
$M \to M - M_c^W$ is a good
fit to the data.  Finally we compute the  
strange quark mass $m_{\overline{\rm MS}}$(2 GeV) at three lattice
spacings.  Within our errors, the results are scale independent
so we take a weighted average giving 
$m_{\overline{\rm MS}}(2 ~{\rm GeV}) ~=~ 95(26)$ MeV,
where systematic uncertainties (except for quenching effects)
have been added in quadrature with the statistical error.  

We intend to perform a larger scale calculation which will
include the vector meson spectrum.  This will allow us to
calculate the ``average'' light quark mass $(m_u + m_d)/2$ 
together with the strange quark mass.  
In addition we will be able to estimate the systematic error
due to setting the lattice scale from $f_\pi$ vs.\ $M_\rho$.
A higher statistics study will be able to sensitively test 
for scaling violations, and possibly give a value
for the light quark masses which is comparable in precision
to other lattice calculations.

Future prospects for this formulation are promising.
Domain wall fermions have an advantage over Wilson fermions,
improved or not, in that they have chiral symmetry which is broken
only by the explicit mass $m$ coupling the boundaries -- the mixing of
the two modes in between the boundaries can be made negligible
compared to $m$.  Consequently there is no mixing between operators of
different chirality and the quark mass is protected from additive
renormalization.  A lattice discretization which respects the axial
symmetries of continuum QCD has an excellent chance to improve 
calculations of matrix elements involving light hadrons.


\section*{Acknowledgements}

We are grateful for interesting and useful discussions with 
Michael Creutz, Chris Dawson,
Robert Mawhinney, Yigal Shamir, Shigemi Ohta,
and the lattice group at Columbia University.
Simulations were performed on the Cray T3E's at the National
Energy Research Supercomputer Center.

\appendix

\section{Diagonalization}
\label{sec:diag}

\subsection{Tree level}
\label{sec:diag0}

In this Appendix we discuss the spectrum of the domain wall
Dirac operator at zero momentum at tree level.  Although 
this work appears in Ref.\ \cite{ref:SHAMIR}, 
a treatment here is useful to establish our notation, 
which is similar to Ref.~\cite{ref:AOKI}.

Given the mass matrix $\Omega_0$, defined in Eqn.~(\ref{eq:omega0}),
we wish to solve the eigenvalue equation
\beq 
\Big(\Omega_0 \Omega_0^\dagger\Big)_{s,s'} \phi^{(i)}_{s'} ~=~ 
(\lambda_0^{(i)})^2 \phi^{(i)}_s,
\eeq
where the index $i\in[1,N_s]$ labels the eigenstates.
The general equation to be satisfied is
\beq
\Big( 1 + b_0^2 - \lambda_0^2 \Big) \phi_s - b_0\Big( \phi_{s+1} + \phi_{s-1}
\Big) ~=~ 0.
\label{eq:gen0}
\eeq
If $0 < b_0 < 1$ then the general solutions are of the form
\beq
\phi_s ~=~ A e^{\alpha_0 s} + B e^{-\alpha_0 s}
\eeq
where $\alpha_0$ is defined through
\beq
\cosh \alpha_0 ~=~ {1+b_0^2 - (\lambda_0^{(i)})^2\over 2 |b_0|}.
\eeq
If $-1 < b_0 < 0$ then the general solutions are of the form
\beq
\phi_s ~=~ (-1)^s\Big(A e^{\alpha_0 s} + B e^{-\alpha_0 s}\Big).
\eeq
For $(\lambda_0^{(i)})^2 < (1-b_0)^2$ or $(\lambda_0^{(i)})^2 > (1+b_0)^2$,
$\alpha$ becomes imaginary and 
\beq
\cos i\alpha_0 ~=~ {1+b_0^2 - (\lambda_0^{(i)})^2\over 2 b_0}.
\eeq

Let us concentrate on the first case of exponential damping.
We must apply the boundary conditions
\bea
\Big(b_0^2 + a^2 m^2 - (\lambda_0^{(1)})^2\Big) \phi_1 - b_0 ~\phi_2 
+ am ~b_0 ~\phi_{N_s}
& = & 0 \\
\Big(1 + b_0^2 - (\lambda_0^{(1)})^2\Big) \phi_{N_s} - b_0 ~\phi_{N_s-1}
+ am ~b_0 ~\phi_1 & = & 0.
\eea
In order to make these conditions consistent with the general
equation (\ref{eq:gen0}), $\phi_{s=0}$ and $\phi_{s=N_s+1}$
must satisfy
\bea
-b_0 ~\phi_0 ~+ ~\phi_1 & = & (am)^2~\phi_1 - am ~b_0 ~\phi_{N_s}
\\
-b_0 ~\phi_{N_s+1} & = & am ~b_0 ~\phi_1.
\eea
As has been found previously \cite{ref:SHAMIR,ref:AOKI,ref:VRANAS},
the eigenvalue is
\bea
(\lambda^{(1)})^2 &=& (am)^2 (1 - b_0^2)^2 ~+~ O\Big((am)^4\Big)
~+~ O\Big( b_0^{N_s} \Big) \nonumber \\
&=& (am)^2 M^2 (2-M)^2 ~+~ O\Big((am)^4\Big) ~+~ O\Big( (1-M)^{N_s}\Big).
\eea
The corresponding normalized eigenvector is
\bea
\phi_s^{(1)} &=& \sqrt{1 - b_0^2} ~e^{-\alpha_0(s-1)} \nonumber \\
&=& \sqrt{M(2-M)} ~e^{-\alpha_0(s-1)}.
\label{eq:tree_basis}
\eea
The eigenvectors corresponding to the heavier modes can be
decomposed into a basis of sine functions
\beq
\phi_s^{(i)} ~=~ \sqrt{ { 2\over N_s}} ~\sin \bigg( {\pi (i-1)\over N_s}
\Big[ N_s + 1 - s\Big]\bigg); \hspace{0.3cm}{ \rm for}~ i\ne 1
\eeq

Note that if $b_0 < 0$ then
\beq 
\phi_s^{(1)} ~=~ (-1)^s \sqrt{1 - b_0^2} ~e^{-\alpha_0(s-1)}
\eeq
while the other eigenvectors and all the eigenvalues
are unchanged.  Also, the eigenvalues and eigenvectors are the same 
for $\Omega_0^\dagger \Omega_0$ with the 
substitution $s \rightarrow N_s+1-s$.
Let us define unitary matrices which diagonalize $\Omega_0 \Omega_0^\dagger$
as
\beq
U^{(0)}_{s,s'} ~\equiv \phi^{(s)}_{s'}
\eeq
and which diagonalize $\Omega_0^\dagger \Omega_0$ as
\beq
V^{(0)}_{s,s'} ~\equiv \phi^{(s)}_{N_s+1-s'}.
\eeq


\subsection{One--loop level}
\label{sec:diag1}

As in Appendix~\ref{sec:diag0}, we want to derive an effective action
for the lightest eigenstate.  In general this involves
computing one--loop corrections to the matrices $U$ and 
$V$ which diagonalize the square of the mass matrix, $\Omega$.
 However, it has already been shown \cite{ref:AOKI,ref:KNY}
that the light mode is stable under radiative corrections.
Following Ref.~\cite{ref:AOKI}, let us write the mass matrix as
\beq
\Omega ~=~ W^- ~+~ g^2 \Omega_1
 ~\equiv~ W^- ~+~ {g^2 C_F\over 16\pi^2} \Big(\Sigma^{(0),-} ~+~ 
 am\Sigma^{(2),-} \Big),
\eeq
where the $\Sigma^{(j)}$ are given in Eqns. (\ref{eq:sigma0})
-- (\ref{eq:sigma2}) and $\Sigma^{(j),\pm} \equiv \Sigma^{(j)}P_\pm$.
Then, to diagonalize $\Omega\Omega^\dagger$ to $O(g^2)$, one
must compute the corrections to $U$ and $V$:
\bea
U^{(0)} &\rightarrow& U \equiv (1+ g^2 U^{(1)})U^{(0)} \nonumber \\
V^{(0)} &\rightarrow& V \equiv (1+ g^2 V^{(1)})V^{(0)},
\eea
If we write the one--loop eigenvalue equations as
\beq
\Big( U\Omega\Omega^\dagger U^\dagger\Big)_{s,s'} ~=~ 
\Big[(\lambda_0^{(s)})^2 + g^2 (\lambda_1^{(s)})^2\Big]\delta_{s,s'},
\eeq
and
\beq
\Big( V\Omega^\dagger \Omega V^\dagger\Big)_{s,s'} ~=~ 
\Big[(\lambda_0^{(s)})^2 + g^2 (\lambda_1^{(s)})^2\Big]\delta_{s,s'},
\eeq
then it can be shown that \cite{ref:AOKI}
\bea
U^{(1)}_{s,s'} &=& \bigg[ \lambda_0^{(s)}(U^{(0)} \Omega_1 V^{(0)}{}^\dagger)_{s,s'}
+ \lambda_0^{(s')}(U^{(0)} \Omega_1 V^{(0)}{}^\dagger)_{s',s} \bigg] \bigg/ 
\bigg[ (\lambda_0^{(s)})^2 - (\lambda_0^{(s')})^2 \bigg] \nonumber \\
V^{(1)}_{s,s'} &=& \bigg[ (U^{(0)} \Omega_1 V^{(0)}{}^\dagger)_{s,s'}\lambda_0^{(s')}
+ (U^{(0)} \Omega_1 V^{(0)}{}^\dagger)_{s',s}\lambda_0^{(s)} \bigg] \bigg/
\bigg[ (\lambda_0^{(s)})^2 - (\lambda_0^{(s')})^2 \bigg] \nonumber
\eea
for $s\ne s'$, and
\beq
U^{(1)}_{s,s} ~=~ V^{(1)}_{s,s} ~=~ 0,
\eeq
and, most importantly,
\beq
(\lambda_1^{(s)})^2 ~=~ 2(U^{(0)} \Omega_1 V^{(0)}{}^\dagger)_{s,s}
\lambda_0^{(s)}.
\eeq
Therefore, in order to compute the quark mass to one loop,
we need $(U^{(0)}\Omega_1 V^{(0)}{}^\dagger)_{1,1}$.  Aoki and Taniguchi
\cite{ref:AOKI} have shown that $U^{(0)}\Sigma_{0,-}V^{(0)}{}^\dagger$ is
negligibly small, and in Appendix~\ref{sec:details} we compute 
$U^{(0)}\Sigma^{(2),-}V^{(0)}{}^\dagger\Big|_{1,1}$ and
$V^{(0)}\Sigma^{(2),+}U^{(0)}{}^\dagger\Big|_{1,1}$.

\section{Perturbative calculation -- details}
\label{sec:details}

\subsection{Feynman rules}
\label{sec:rules}

The Feynman rules are similar to those for Wilson fermions
plus plaquette--action gluons~\cite{ref:KAWAI}. 

In the gauge sector, we use the usual four dimensional plaquette
action:
\beq
S_g ~=~ \beta \sum_{\rm plaq} {1\over3}~{\rm Re~Tr~}(1 - U_{\rm plaq}).
\eeq
In the five dimensional picture there are simply $N_s$ copies
of the gauge field, and in the flavor interpretation the gluons
are simply flavorless.

The propagator of a gluon (in Feynman gauge) with momentum $q$ is given by
\beq
D_{\mu\nu}(q) ~=~ {\delta_{\mu\nu}\over 4 \sum_\rho \sin^2(aq_\rho/2)}
~\equiv~ {\delta_{\mu\nu}\over \hat{q}^2}.
\eeq
The one-gluon -- fermion vertex with incoming (outgoing) momentum
$p_1$ ($p_2$) is $g t_n v_\mu(q)$, where
\beq
v_\mu(q) ~=~ \Big( i\gamma_\mu \cos aq_\mu - \sin aq_\mu \Big),
\eeq
$q = (p_1 + p_2)/2$, and $t_n$ is one of the eight generators
of $SU(3)$.
The two-gluon -- fermion vertex with gluon indices $n,\mu$ and $n,\mu'$
and fermion momenta as above is $(g^2/2)\{t_n,t_{n'}\}\delta_{\mu,\mu'}$,
with
\beq
w_\mu(q) ~=~ -\Big( i\gamma_\mu \sin aq_\mu + \cos aq_\mu \Big).
\eeq

The massive fermion propagator for the boundary wall variant used
in this work was derived in Ref.~\cite{ref:SHAMIR} and also
Ref.~\cite{ref:AOKI} whose notation we adopt.  The tree--level
$G^R$ and $G^L$, as defined in Eqn.~(\ref{eq:glgr}), are given
by
\beq
G^R_{s,s'} ~=~ G^0_{s,s'} + A_{++}e^{\alpha(s+s')} + 2A_{+-}\cosh\alpha(s-s')
+ A_{--}e^{-\alpha(s+s')}
\eeq
and
\beq
G^L_{s,s'} ~=~ G^0_{s,s'} + B_{++}e^{\alpha(s+s')} + 2B_{+-}\cosh\alpha(s-s')
+ B_{--}e^{-\alpha(s+s')}.
\eeq
The inhomogeneous part of the solution is
\beq
G^0_{s,s'} ~=~ A \cosh\Big( \alpha(N_s -|s-s'|)\Big),
\eeq
and the constants defined implicitly above are
\bea
A & = & {1\over 4 b\sinh\alpha ~ \sinh(\alpha N_s)} \\
A_{++} & = & (e^{-2\alpha N_s} - 1)(1 - be^e{-\alpha})(1-m^2)A/F \\
A_{+-} & = & 2b\sinh\alpha(1+2m\cosh(\alpha N_s) + m^2)A/F \\
A_{--} & = & (1-e^{2\alpha N_s})(1-be^\alpha)(1-m^2)A/F \\
B_{++} & = & (e^{-2\alpha N_s} - 1)e^{-\alpha}(e^{-\alpha}-b)(1-m^2)A/F \\
B_{+-} & = & 2b\sinh\alpha(1+2m\cosh(\alpha N_s) + m^2)A/F \\
B_{--} & = & (1-e^{2\alpha N_s})e^\alpha(e^\alpha-b)(1-m^2)A/F
\eea
where
\beq
F = e^{\alpha N_s}\Big[1-be^\alpha +m^2(be^{-\alpha}-1)\Big]
- 4bm\sinh\alpha + e^{-\alpha N_s}\Big[be^{-\alpha} - 1 +m^2(1-be^\alpha)
\Big].
\eeq

\subsection{Tadpole diagram}
\label{sec:tadpole}

The tadpole diagram (Fig.~\ref{fig:loops}a) is simple to compute.
Since it has no  fermion propagator in the loop, it is equivalent
to the case for Wilson fermions:
\bea
\Sigma^{\rm tad}_{s,s'}(p) &=& {16\pi^2\delta_{s,s'}\over a} 
\sum_\mu \loopint{l} w_\mu(ap) D_{\mu\nu}(l) \nonumber \\
&=& - { \delta_{s,s'}\over 2}\Big( i\pslash{p} + {4\over a} \Big) T,
\eea
where $T$ is the finite integral
\beq
T ~=~ 16\pi^2\loopint{l} {1\over 4\sum_\mu\sin^2(l_\mu/2)} ~=~ 24.4.
\eeq

\subsection{Taylor expansion}
\label{sec:taylor}

The calculation of the half--circle graph (Fig.~\ref{fig:loops}b)
with domain wall fermions parallels the same calculation with Wilson 
fermions~\cite{ref:BSD}.
The contribution of the half--circle graph to the fermion
self--energy is
\bea
\Sigma^{\rm h-c}_{s,s'}(p,m) &=& {16\pi^2\over a} \loopint{l}
\sum_{\mu,\nu} v_\mu\Big({l+ap\over 2}\Big) S^F_{s,s'}(l,am) 
D_{\mu\nu}(ap-l) v_\nu\Big({l+ap\over2}\Big).
\label{eq:h-c} \\
&\equiv& {16\pi^2\over a} \loopint{l} E(l,ap,am).
\eea
The second line above defines the integrand $E$.

A simple Taylor expansion of Eqn.~(\ref{eq:h-c}) about zero lattice
spacing would not be valid due to the logarithmic divergence
of the integral.  That is, the coefficients of the power series
in $a$ would have a $\ln a$ dependence, which must first be
separated before expanding the Taylor series.
We subtract and then add back a similar half--circle graph built
from continuum--like Feynman rules designed to have the same
infrared behavior as the present rules.
\bea
\Sigma^{\rm IR}_{s,s'}(p,m) &=& {16\pi^2\over a} \loopint{l}
~\Theta(\pi^2 - l^2)
~{(i\gamma_\mu)~S^{\rm IR}_{s,s'}(l,am)~(i\gamma_\mu)\over (ap-l)^2} \\
&\equiv& {16\pi^2\over a} \loopint{l} E^{\rm IR}(l,ap,am).
\eea
The step function $\Theta(\pi^2 - l^2)$ makes the integral
spherically symmetric and therefore easier to evaluate.
We will specify the continuum--like fermion propagator $S^{\rm IR}$
in the next Section.

Now we can write the half-circle graph in terms of a Taylor
expansion of $(\Sigma - \Sigma^{\rm IR})$
\bea
\Sigma^{\rm h-c}(p,m) & = & {16\pi^2\over a}\int_l
~\Theta(\pi^2 - l^2) E^{\rm IR}(l,ap,am) \nonumber \\
& + & {1\over a}\int_l
\Big[ E(l,ap,am) - \Theta(\pi^2 - l^2)E^{\rm IR}(l,ap,am) \Big]
\nonumber \\
& = & {16\pi^2\over a}\int_l ~\Theta(\pi^2 - l^2) \Big[ E^{\rm IR}(l,ap,am)
- E^{\rm IR}(l,0,0) \Big] + {1\over a}\int_l E(l,0,0)
\nonumber \\
& + & \int_l {d\over da}\Big[ E(l,ap,am) - \Theta(\pi^2 - l^2)
E^{\rm IR}(l,ap,am) \Big]
\Big|_{a=0} ~+~ O(a).
\label{eq:taylor}
\eea
We use a shorthand notation for the loop integral, namely
$\int_l \equiv \int d^4 l/(2\pi)^4$.  Eqn.~(\ref{eq:taylor})
is the ``master'' equation for computing the half--circle graph:
in the remainder of this Appendix, we compute the various terms
appearing within it.

\subsection{Infrared terms}
\label{sec:ir}

In this Section, we compute $E^{\rm IR}$ and $dE^{\rm IR}/da$.
The IR ($a \to 0$) limit of the fermion propagator is
\beq
S^{\rm IR}(l,am) ~=~ {1\over l^2 + a^2 (m_q^{(0)})^2}
\sum_{+,-} ( -i\lslash - a m \delta_\pm + \Delta_\pm) C_\pm P_\pm,
\label{eq:irprop}
\eeq
where $\delta_+ = \delta_{s,1}\delta_{s',N_s}$,
$\delta_- = \delta_{s,N_s}\delta_{s',1}$, $\Delta_\pm = \delta_{s\mp1,s'}
- b_0\delta_{s,s'}$, and 
\bea
C_+ &=& (1-b_0^2) b_0^{2N_s-s-s'} + am_q^{(0)}b_0^{N_s}
\Big( b_0^{s-s'} + b_0^{-s+s'} \Big) \\
C_- &=& (1-b_0^2) b_0^{s+s'-2} + am_q^{(0)}b_0^{N_s}
\Big( b_0^{s-s'} + b_0^{-s+s'} \Big) .
\eea
Note also the presence of the tree--level quark mass
$m_q^{(0)} = mM(2-M)$ which was defined above in Eqn.~(\ref{eq:mq0}).

Let us first compute the integral of $dE^{\rm IR}/da$.
Since the IR vertices are just $i\gamma_\mu$, there will
be just two terms, one with $dS^{\rm IR}(l,am)/da$
and the other with $d(ap-l)^{-2}/da$.  The result for the
first term is
\beq
\int_l \Theta \sum_\mu { (i\gamma_\mu) S^{\rm IR}(l,am)
(-2p)\cdot(ap-l)(i\gamma_\mu) \over (ap-l)^4} \bigg|_{a=0}
~=~ \int_l \Theta \sum_{+,-} {-i\pslash{p} C^0_\mp P_\pm
\over l^4},
\label{eq:dEirda1}
\eeq
where $C^0 \equiv C|_{a=0}$, and $\Theta = \Theta(\pi^2 - l^2)$.
Note the $\Delta_\pm$ term in the fermion propagator is anti-symmetric
in $l$ and so vanishes upon integration above.  The second term
comes from taking the derivative of $S^{\rm IR}$:
\bea
{d\over da}S^{\rm IR}(l,am)\Big|_{a=0} & = &
-S^{\rm IR}(l,0) \bigg[ {d\over da} 
\Big(S^{\rm IR}(l,am)\Big)^{-1} \bigg] \bigg|_{a=0} S^{\rm IR}(l,0)
\nonumber \\
&=&
\sum_{+,-} {m\over l^4}\bigg[ (-i\lslash + \Delta_\pm) 
~ C^0_\pm P_\pm \delta_\pm ~ (-i\lslash C^0_\pm + \Delta_\pm C^0_\mp) P_\pm
\bigg]
\eea
After some algebra, we have
\bea
\int_l {1\over l^2} \bigg[ (i\gamma_\mu){d\over da}S^{\rm IR}(i\gamma_\mu)
\bigg]_{s,s'}\bigg|_{a=0} &=& -4(16\pi^2) m_q^{(0)}  \int_l 
\Theta { 1-b_0^2\over l^4} \Big[ b_0^{N_s-s+s'-1}P_+
\nonumber \\ 
&  & -b_0^{N_s +s -s' -1} P_- \Big].
\label{eq:dEirda2}
\eea

The integration of the $E^{\rm IR}(l,ap,am) - E^{\rm IR}(l,0,0)$ term
of Eqn.~(\ref{eq:taylor})
is quite similar to that for Wilson fermions (see e.g.~\cite{ref:BSD}).
We delay writing down the answer, since it simplifies greatly
upon diagonalization in flavor space.

\subsection{Finite terms}
\label{sec:finite}

Let's first look at the numerator of $E(l,ap,am)$.  For the 
time being we suppress the indices $s,s'$.
\bea
{\cal N} &\equiv& \sum_\mu v_\mu((l+ap)/2) S^F(l,am) v_\mu((l+ap)/2)
\nonumber \\
& = &\sum_\mu \Big[ i\gamma_\mu \cos{1\over2}(l+ap)_\mu - 
\sin{1\over2}(l+ap)_\mu \Big] 
\Big[(-i\bar{\lslash}  + W^-)G^R P_+
+ (-i\bar{\lslash} + W^+)G^L P_- \Big]
\nonumber \\
& \times & \Big[ i\gamma_\mu \cos{1\over2}(l+ap)_\mu - 
\sin{1\over2}(l+ap)_\mu \Big]
\eea
Multiplying the factors of ${\cal N}$ gives
\bea
{\cal N} = & - & \sum_\mu \Big(-i\gamma_\mu~\bar{\lslash}~\gamma_\mu ~+~W^-\Big)
~G^R P_- ~\Big( \cos^2{1\over2}(l+ap)_\mu\Big) \nonumber \\
&-& \sum_\mu\Big(-i\gamma_\mu~\bar{\lslash}~\gamma_\mu ~+~ W^+\Big) ~G^L P_+
~\Big( \cos^2{1\over2}(l+ap)_\mu\Big) \nonumber \\
&+& \Big(-i\bar{\lslash} ~+~ W^-\Big) ~G^R P_+
~\Big( \sum_\mu \sin^2{1\over2}(l+ap)_\mu\Big) \nonumber \\
&+& \Big(-i\bar{\lslash} ~+~ W^+\Big) ~G^L P_-
~\Big( \sum_\mu \sin^2{1\over2}(l+ap)_\mu\Big) \nonumber \\
&-& \sum_\mu i\gamma_\mu\Big(-i\bar{\lslash} ~+~ W^-\Big) ~G^R P_+
~\Big( {1\over2}\sin(l+ap)_\mu \Big) \nonumber \\
&-& \sum_\mu \Big(-i\bar{\lslash} ~+~ W^-\Big)i\gamma_\mu ~G^R P_-
~\Big( {1\over2}\sin(l+ap)_\mu \Big) \nonumber \\
&-& \sum_\mu i\gamma_\mu\Big(-i\bar{\lslash} ~+~ W^+\Big) ~G^L P_-
~\Big( {1\over2}\sin(l+ap)_\mu \Big) \nonumber \\
&-& \sum_\mu \Big(-i\bar{\lslash} ~+~ W^+\Big)i\gamma_\mu ~G^L P_+
~\Big( {1\over2}\sin(l+ap)_\mu \Big),
\label{eq:num1}
\eea
where, as usual, $\bar{\lslash} \equiv \sum_\nu \gamma_\nu \bar{l}_\nu$.
To compute $\int_l E(l,0,0)$ we divide Eqn.~(\ref{eq:num1}) by $\hat{l}^2$
and integrate.
Since the integration region is symmetric in $l_\mu$, the terms odd in
$l_\mu$ in (\ref{eq:num1}) vanish upon integration.  The result is
given by
\bea
\int_l E(l,0,0) &=& \int_l ~{1\over\hat{l}^2} 
\bigg\{ \sum_\mu \sin^2{l_\mu\over2} \Big[ (W_0^- G^R)_{s,s'}P_+ 
+ (W_0^+G^L)_{s,s'}P_- \Big]\nonumber \\
& - & \sum_\mu \cos^2{l_\mu\over2} \Big[(W_0^+ G^L)_{s,s'} P_+
+ (W_0^-G^R)_{s,s'}P_- \Big] 
- {\bar{l}^2\over2}\Big( G^R_{s,s'} + G^L_{s,s'} \Big) \bigg\}
\eea
where $W_0^\pm \equiv W^\pm|_{am=0}$.

Next we compute 
\beq
{d\over da} E(l,ap,am) \bigg|_{a=0} ~\equiv~
i\pslash{p}{\cal E}^1 ~+~ m{\cal E}^2,
\eeq
where
\bea
i\pslash{p}{\cal E}^1 & = & \sum_\mu
2v_\mu((l+ap)/2)S^F(l,am){d\over da}\bigg(v_\mu((l+ap)/2)\bigg)
\widehat{(l-ap)}^{-2} \nonumber \\
&  & +~{\cal N}{d\over da}\bigg(\widehat{(l-ap)}^{-2}\bigg) \nonumber \\
{\rm and}~~m{\cal E}^2 & = & v_\mu((l+ap)/2)~ {d S^F\over da}~ v_\mu((l+ap)/2)
~\widehat{(l-ap)}^{-2}.
\label{eq:cale}
\eea
Note that $v_\mu S^F (dv_\mu/da)$ 
is the same as Eqn.~(\ref{eq:num1}) with the replacements
\bea
\cos^2{1\over2}(l+ap)_\mu & \rightarrow & -{p_\mu\over2}\sin(l+ap)_\mu
\nonumber \\
\sin^2{1\over2}(l+ap)_\mu & \rightarrow & {p_\mu\over2}\sin(l+ap)_\mu
\\
{1\over2}\sin(l+ap)_\mu & \rightarrow & {p_\mu\over2}\cos(l+ap)_\mu.
\nonumber
\eea
After some algebra, the result is ${\cal E}^1_{s,s'} = 
{\cal E}^{1,+}_{s,s'}P_+ + {\cal E}^{1,-}_{s,s'}P_-$, where
\bea
\int_l {\cal E}^{1,+}_{s,s'} &=& -\int_l \bigg\{ {1\over 8\hat{l}^2} \bigg[
\bar{l}^2 \Big(G^R_{s,s'} + G^L_{s,s'}\Big) + \sum_\mu \cos l_\mu
\Big( (W_0^- G^R)_{s,s'} + (W_0^+ G^L)_{s,s'}\Big)\bigg] 
\nonumber \\
& + & {1\over 2\hat{l}^4} \bigg[ \bar{l}^2 \Big(\sum_\mu \sin^2{l\mu\over2}
\Big) G^R_{s,s'} - \sum_\mu\bar{l}^2_\mu\Big( 2\cos^2{l_\mu\over2}
- \sum_\nu \cos^2{l_\nu\over2}\Big) G^L_{s,s'} \nonumber \\
& + &{\bar{l}^2\over2}\Big( (W_0^- G^R)_{s,s'} + (W_0^+ G^L)_{s,s'}\Big)
\bigg] \bigg\}
\label{eq:e1p}
\eea
and
\bea
\int_l {\cal E}^{1,-}_{s,s'} &=& -\int_l \bigg\{ {1\over 8\hat{l}^2} \bigg[
\bar{l}^2 \Big(G^R_{s,s'} + G^L_{s,s'}\Big) + \sum_\mu \cos l_\mu
\Big( (W_0^- G^R)_{s,s'} + (W_0^+ G^L)_{s,s'}\Big)\bigg] 
\nonumber \\
& + & {1\over 2\hat{l}^4} \bigg[ \bar{l}^2 \Big(\sum_\mu \sin^2{l\mu\over2}
\Big) G^L_{s,s'} - \sum_\mu\bar{l}^2_\mu\Big( 2\cos^2{l_\mu\over2}
- \sum_\nu \cos^2{l_\nu\over2}\Big) G^R_{s,s'} \nonumber \\
& + &{\bar{l}^2\over2}\Big( (W_0^- G^R)_{s,s'} + (W_0^+ G^L)_{s,s'}\Big)
\bigg] \bigg\}.
\label{eq:e1m}
\eea
For the second term in Eqn.~(\ref{eq:cale}), the calculation
is simplified through the use of the identity
\bea
{d\over da} S_{s,s'}(l,am)\bigg|_{a=0} &=& -S_{s,s''}(l,0)
\bigg[ {d\over da} D_{s'',s'''}(l,am) \bigg] \bigg|_{a=0} S_{s''',s'}(l,0)
\nonumber \\
&=& S_{s,s''}(l,0)\Big[ m \Big( \delta_{s'',N_s}\delta_{s''',1} P_+
~+~ \delta_{s'',1}\delta_{s''',N_s} P_- \Big) \Big] S_{s''',s'}(l,0).
\eea
The term generated by taking this derivative is proportional, therefore,
to the mass $m$.  We express the coefficient by
\beq
{\cal E}^{2}_{s,s'} ~=~  \sum_\mu ~{1\over\hat{l}^2} 
~v_\mu\Big({l_\mu\over 2}\Big)
~\Big[ S_{s,N_s}P_+ S_{1,s'} ~+~ S_{s,1}P_-S_{N_s,s'} \Big]
~v_\mu\Big({l_\mu\over 2}\Big) .
\eeq
Algebraic manipulation similar to above gives
 ${\cal E}^{2}_{s,s'} = {\cal E}^{2,+}_{s,s'} P_+ 
 + {\cal E}^{2,-}_{s,s'} P_-$ with
\bea
\int_l {\cal E}^{2,+}_{s,s'}& = &\int_l {1\over\hat{l}^2} \bigg\{ 
\bar{l}^2\sum_\mu \cos^2{l_\mu\over2} G^R_{s,N_s}G^L_{1,s'}
~-~ \bar{l}^2\sum_\mu \sin^2{l_\mu\over2} G^L_{s,1}G^R_{N_s,s'}
\nonumber \\
& + & \sum_\mu \sin^2{l_\mu\over2}(W_0^-G^R)_{s,N_s}(W_0^-G^R)_{1,s'}
~-~ \sum_\mu \cos^2{l_\mu\over2}(W_0^+G^L)_{s,1}(W_0^+G^L)_{N_s,s'}
\nonumber \\
& - & {\bar{l}^2\over2} \bigg[ G^R_{s,N_s}(W_0^-G^R)_{1,s'}
~+~ (W_0^-G^R)_{s,N_s}G^L_{1,s'} ~+~ G^L_{s,1}(W_0^+G^L)_{N_s,s'}
\nonumber \\
& + & (W_0^+G^L)_{s,1}G^R_{N_s,s'} \bigg] \bigg\}
\label{eq:e2p}
\eea
and
\bea
\int_l {\cal E}^{2,-}_{s,s'}& = & \int_l {1\over\hat{l}^2} \bigg\{ 
\bar{l}^2\sum_\mu \cos^2{l_\mu\over2} G^L_{s,1}G^R_{N_s,s'}
~-~ \bar{l}^2\sum_\mu \sin^2{l_\mu\over2} G^R_{s,N_s}G^L_{1,s'}
\nonumber \\
& + & \sum_\mu \sin^2{l_\mu\over2}(W_0^+G^L)_{s,1}(W_0^+G^L)_{N_s,s'}
~-~ \sum_\mu \cos^2{l_\mu\over2}(W_0^-G^R)_{s,N_s}(W_0^-G^R)_{1,s'}
\nonumber \\
& - & {\bar{l}^2\over2} \bigg[ G^R_{s,N_s}(W_0^-G^R)_{1,s'}
~+~ (W_0^-G^R)_{s,N_s}G^L_{1,s'} ~+~ G^L_{s,1}(W_0^+G^L)_{N_s,s'}
\nonumber \\
& + & (W_0^+G^L)_{s,1}G^R_{N_s,s'} \bigg] \bigg\}
\label{eq:e2m}
\eea

\subsection{Diagonalization of the self--energy}
\label{sec:diag_sigma}

Returning to the calculation of 
$E^{\rm IR}(l,ap,am) - E^{\rm IR}(l,0,0)$, we diagonalize
the IR fermion propagator.  We refer the reader to 
Appendix~\ref{sec:diag} for the explanation of the matrices
which diagonalize the mass matrix.  In the $a\to 0$ limit,
the propagator for light mode is
\beq
S^{\rm IR}_{1,1}(l,am) ~=~ {1\over l^2 + a^2 (m_q^{(0)})^2}
\sum_{+,-} ( -i\lslash + a m ),
\eeq
since the pieces of the full $5d$ propagator (\ref{eq:irprop}) 
transform as follows in the diagonal basis:
\bea
UC_+U^\dagger &=& VC_-V^\dagger ~=~ 1 + O(N_s b_0^{N_s}) \\
V\Delta_-C_-U^\dagger &=& U\Delta_+C_+V^\dagger ~=~ O(N_s^2 b_0^{2N_s}) \\
V\delta_-C_-U^\dagger &=& U\delta_+C_+V^\dagger ~=~ b_0 (1-b_0^2) am.
\eea
Then the calculation of the $E_{\rm IR}(l,ap,am) - E_{\rm IR}(l,0,0)$
term
proceeds identically to that in the case of ordinary four--dimensional
Wilson fermions~\cite{ref:GONZ,ref:BSD}.  The result is
\bea
{16\pi^2\over a}\int_l \Theta(\pi^2 - l^2) \Big[ E_{\rm IR}(l,ap,am) 
&-&  E_{\rm IR}(l,0,0)\Big] ~=~ - \int_0^1 dx ~\nonumber \\
&\times&
\bigg\{ 2i\pslash{p}x\bigg[ \ln\Big({\pi\over az}\Big)^2 - {3\over 2}\bigg]
+ 4m_q^{(0)} \bigg[ \ln\Big({\pi\over az}\Big)^2 -1 \bigg]\bigg\}
\nonumber \\
&\equiv& \bigg(i\pslash{p} \tilde{L}_1 ~+~ m_q^{(0)}\tilde{L}_2\bigg),
\eea
where $z^2 \equiv (1-x)(p^2 x + (m_q^{(0)})^2)$.

Next we compute the derivative term of Eqn.~(\ref{eq:taylor}) for the
light mode.  Recalling the notation from Eqn.~(\ref{eq:sigma1}),
\beq
16\pi^2\int_l {d\over da}\Big[ E(l,ap,am) - \Theta(\pi^2 - l^2)
E^{\rm IR}(l,ap,am) \Big]
\Big|_{a=0} ~\equiv~ i\pslash{p} \tilde{I}_1 ~+~ m_q^{(0)} \tilde{I}_2.
\eeq
Combining ${\cal E}^1$ (Eqns.\ (\ref{eq:e1p}) and (\ref{eq:e1m}))
and its IR counterpart (\ref{eq:dEirda1}) gives
\bea
\tilde{I}_1 &=& 16\pi^2\loopint{l}~ \Bigg\{ {1\over 8 \hat{l}^2} \sum_\mu \bigg[
\sin^2l_\mu (\tilde{G}_R + \tilde{G}_L) + 2 \cos l_\mu (b_0 - b(l))
\tilde{G}_R\Big] \nonumber \\
&+& \sum_\mu {\sin^2 l_\mu\over 2 \hat{l}^4}\Big[(b_0-b(l))\tilde{G}_R
- \Big( 2\cos^2{l_\mu\over2} - \sum_\nu \cos^2{l_\nu\over2}\Big)\tilde{G}_L
+ \hat{l}^2 \tilde{G}_R \Big] \Bigg\}
\nonumber \\
& - & 16\pi^2\loopint{l} ~\Theta(\pi^2 - l^2) ~{1\over l^4}
\eea
where 
\bea
\tilde{G}_R(l) &\equiv& \sum_{s,s'} (U^{(0)})_{1,s} G^R_{s,s'}(l)
 (U^{(0)}{}^\dagger)_{s',1}
~=~ \sum_{s,s'} (V^{(0)})_{1,s} G^L_{s,s'}(l) (V^{(0)}{}^\dagger)_{s',1}
\nonumber \\
& = & {1\over 2b(l)\sinh\alpha(l)}\bigg[ {(b_0^{-1}-e^{-\alpha(l)}) 
- (b_0-e^{\alpha(l)})
\over (b_0^{-1}-e^{-\alpha(l)}) + (b_0-e^{\alpha(l)})} 
~-~ {1-b_0^2 \over (e^{\alpha(l)} - b_0)^2} \bigg]
\eea
and
\bea
\tilde{G}_L(l) &\equiv& \sum_{s,s'} (U^{(0)})_{1,s} G^L_{s,s'}(l)
 (U^{(0)}{}^\dagger)_{s',1}
~=~ \sum_{s,s'} (V^{(0)})_{1,s} G^R_{s,s'}(l) (V^{(0)}{}^\dagger)_{s',1}
\nonumber \\
& = & {1\over 2b(l)\sinh\alpha(l)}\bigg[ {(b_0^{-1}-e^{-\alpha(l)}) 
- (b_0-e^{\alpha(l)})
\over (b_0^{-1}-e^{-\alpha(l)}) + (b_0-e^{\alpha(l)})} 
~-~ {1-b_0^2\over
(e^{\alpha(l)} - b_0)^2}{e^{\alpha(l)} - b(l) \over e^{-\alpha(l)}-b(l)} \bigg]
\eea
Likewise, combining ${\cal E}^2$ (Eqns.\ (\ref{eq:e2p}) and (\ref{eq:e2m}))
and its IR counterpart (\ref{eq:dEirda2}) gives
\bea
\label{eq:i2long}
\tilde{I}_2 & = & {16\pi^2\over 1-b_0^2}\loopint{l}~ {1\over2\hat{l}^2} \Bigg\{
-\bar{l}^2 \sum_\mu \cos^2{l_\mu\over2}
\bigg[ V_{1,s}G^R_{s,N_s}G^L_{1,s'}U^\dagger{}_{s',1} 
+ U_{1,s}G^L_{s,1}G^R_{N_s,s'}V^\dagger{}_{s',1}\bigg] \nonumber \\
& + & \bar{l}^2 \sum_\mu \sin^2{l_\mu\over2}
\bigg[ V_{1,s}G^L_{s,1}G^R_{N_s,s'}U^\dagger{}_{s',N_s} 
+ U_{1,s}G^R_{s,N_s}G^L_{1,s'}V^\dagger{}_{s',N_s}\bigg] \nonumber \\
& + & {\bar{l}^2\over2}\bigg[V_{1,s}G^R_{s,N_s}\Big(W^-G^R\Big)_{1,s'}
U^\dagger{}_{s',1} + U_{1,s}G^R_{s,N_s}\Big(W^-G^R\Big)_{1,s'}
V^\dagger{}_{s',1}\bigg] \nonumber \\ 
& + & {\bar{l}^2\over2}\bigg[V_{1,s}\Big(W^-G^R\Big)_{s,N_s}G^L_{1,s'}
U^\dagger{}_{s',1} + U_{1,s}\Big(W^-G^R\Big)_{s,N_s}G^L_{1,s'}
V^\dagger{}_{s',1}\bigg] \nonumber \\
& + & {\bar{l}^2\over2}\bigg[V_{1,s}G^L_{s,1}\Big(W^+G^L\Big)_{N_s,s'}
U^\dagger{}_{s',1} + U_{1,s}G^L_{s,1}\Big(W^+G^L\Big)_{N_s,s'}
V^\dagger{}_{s',1}\bigg] \nonumber \\
& + & {\bar{l}^2\over2}\bigg[V_{1,s}\Big(W^+G^L\Big)_{s,1}G^R_{N_s,s'}
U^\dagger{}_{s',1} + U_{1,s}\Big(W^+G^L\Big)_{s,1}G^R_{N_s,s'}
V^\dagger{}_{s',1}\bigg] \nonumber \\
& + & \sum_\mu \cos^2{l_\mu\over2}\bigg[V_{1,s}\Big(W^+G^L\Big)_{s,1}
\Big(W^+G^L\Big)_{N_s,s'}U^\dagger{}_{s',1} 
\nonumber \\ 
&  & \hspace{1cm} + ~ U_{1,s}\Big(W^-G^R\Big)_{s,N_s}
\Big(W^-G^R\Big)_{1,s'}V^\dagger{}_{s',1}\bigg] \nonumber \\
& - & \sum_\mu \sin^2{l_\mu\over2}\bigg[V_{1,s}\Big(W^-G^R\Big)_{s,N_s}
\Big(W^-G^R\Big)_{1,s'}U^\dagger{}_{s',1} 
\nonumber \\ 
&  & \hspace{1cm} + ~ U_{1,s}\Big(W^+G^L\Big)_{s,1}
\Big(W^+G^L\Big)_{N_s,s'}V^\dagger{}_{s',1}\bigg]\Bigg\} \nonumber \\
& - &  4(16\pi^2)\loopint{l} ~\Theta(\pi^2 - l^2) ~{1\over l^4}
\eea
Summing over internal indices we find that
\bea
\label{eq:GGdiagfirst}
V_{1,s}G^R_{s,N_s}G^L_{1,s'}U^\dagger{}_{s',1} & = &
{1\over(e^{\alpha(l)} - b_0)^2} ~ {e^{2\alpha(l)}
\over(1-e^{\alpha(l)} b(l))^2}
~\equiv~ G_1(l)
\\
U_{1,s}G^L_{s,1}G^R_{N_s,s'}V^\dagger{}_{s',1} & = & G_1(l) \\
V_{1,s}G^R_{s,N_s}\Big(W^-G^R\Big)_{1,s'}U^\dagger{}_{s',1} & = &
\Big(e^{-\alpha(l)} - b(l)\Big) ~G_1(l) \\
V_{1,s}\Big(W^-G^R\Big)_{s,N_s}G^L_{1,s'}U^\dagger{}_{s',1} & = &
\Big(e^{-\alpha(l)} - b(l)\Big) ~G_1(l) \\
U_{1,s}G^L_{s,1}\Big(W^+G^L\Big)_{N_s,s'}V^\dagger{}_{s',1} & = &
\Big(e^{-\alpha(l)} - b(l)\Big) ~G_1(l) \\
U_{1,s}\Big(W^+G^L\Big)_{s,1}G^R_{N_s,s'}V^\dagger{}_{s',1} & = &
\Big(e^{-\alpha(l)} - b(l)\Big) ~G_1(l) \\
V_{1,s}\Big(W^-G^R\Big)_{s,N_s}\Big(W^-G^R\Big)_{1,s'}U^\dagger{}_{s',1}
& = & \Big(e^{-\alpha(l)} - b(l)\Big)^2 ~G_1(l) \\
U_{1,s}\Big(W^+G^L\Big)_{s,1}\Big(W^+G^L\Big)_{N_s,s'}V^\dagger{}_{s',1}
& = & \Big(e^{-\alpha(l)} - b(l)\Big)^2 ~G_1(l).
\label{eq:GGdiaglast}
\eea
All other terms in (\ref{eq:i2long}) vanish as $N_s\to\infty$.
Substituting (\ref{eq:GGdiagfirst})--(\ref{eq:GGdiaglast})
 into (\ref{eq:i2long}) gives
the final result:
\bea
\tilde{I}_2 & = & -16\pi^2\loopint{l}~ {1\over \hat{l}^2} {1\over(e^{\alpha(l)}
 - b_0)^2}~ \Bigg[ \bar{l}^2 \sum_\mu \cos^2{l_\mu\over2}
{1\over(1 - e^{\alpha(l)} b(l))^2} - \bar{l}^2 {e^{-\alpha(l)}
\over(1 - e^{\alpha(l)} b(l))}
 \nonumber \\
& + &  \sum_\mu\sin{l_\mu\over2} ~e^{-2\alpha(l)} \Bigg]
 -  4(16\pi^2)\loopint{l} ~\Theta(\pi^2 - l^2) ~{1\over l^4}
\eea



\end{document}